\documentclass{emulateapj}

\begin{document}
\def\mpch {$h^{-1}$ Mpc} 
\def\kpch {$h^{-1}$ kpc} 
\def\kms {km s$^{-1}$} 
\def\lcdm {$\Lambda$CDM } 
\def\xir {$\xi(r)$}
\def\wprp {$w_p(r_p)$}
\def\xisp {$\xi(r_p,\pi)$}
\def\xis {$\xi(s)$}
\def\rr {$r_0$}
\def\gg {$\gamma$}
\def\etal {et al.}

\def\kt{\tilde{k}}
\def\mpc{{\rm Mpc}}
\def\zmax{Z_{\rm max}}

\title{The DEEP2 Galaxy Redshift Survey: Color and Luminosity
  Dependence of Galaxy Clustering at $z\sim1$}
\author{Alison L. Coil\altaffilmark{1,2}, 
Jeffrey A. Newman\altaffilmark{3},
Darren Croton\altaffilmark{4}, 
Michael C. Cooper\altaffilmark{4}, 
Marc Davis\altaffilmark{4},
S.~M. Faber\altaffilmark{5},
Brian F. Gerke\altaffilmark{4},
David C. Koo\altaffilmark{5},
Nikhil Padmanabhan\altaffilmark{1,3},
Risa H. Wechsler\altaffilmark{6},
Benjamin J. Weiner\altaffilmark{2}
}
\altaffiltext{1}{Hubble Fellow}
\altaffiltext{2}{Steward Observatory, University of Arizona,
Tucson, AZ 85721}
\altaffiltext{3}{Institute for Nuclear and Particle Astrophysics,
  Lawrence Berkeley National Laboratory, Berkeley, CA 94720} 
\altaffiltext{4}{Department of Astronomy, University of California,
Berkeley, CA 94720 -- 3411} 
\altaffiltext{5}{University of California Observatories/Lick
Observatory, Department of Astronomy and Astrophysics, University of
California, Santa Cruz, CA 95064}
\altaffiltext{6}{Kavli Institute for Particle Astrophysics \& Cosmology,
  Physics Department, and Stanford Linear Accelerator Center,
  Stanford University, Stanford, CA 94305}

\begin{abstract}

We present measurements of the color and luminosity dependence of
galaxy clustering at $z\sim1$ in the DEEP2 Galaxy Redshift Survey.
Using volume-limited subsamples in bins of both color and luminosity,
we find that: 1) The clustering dependence is much stronger with color
than with luminosity and is as strong with color at $z\sim1$ as is found locally.
We find no dependence of the clustering amplitude on color for
galaxies on the red sequence, but a significant dependence on color
for galaxies within the blue cloud.  2) For galaxies in the range 
$L/L^*\sim0.7-2$, a stronger large-scale
luminosity dependence is seen for all galaxies than for red and blue
galaxies separately.  The small-scale clustering
amplitude depends significantly on luminosity for blue galaxies, with
brighter samples having a stronger rise on scales $r_p<0.5$ \mpch.  3)
Redder galaxies exhibit stronger small-scale redshift-space
distortions (``fingers of god''), and both red and blue populations
show large-scale distortions in \xisp \ due to coherent infall.  4)
While the clustering length, \rr, increases smoothly with galaxy color
(in narrow bins), its power-law exponent, $\gamma$, exhibits a sharp
jump from the blue cloud to the red sequence.  The intermediate color
`green' galaxy population likely includes transitional galaxies moving
from the blue cloud to the red sequence; on large scales green
galaxies are as clustered as red galaxies but show infall kinematics
and a small-scale correlation slope akin to the blue galaxy
population.  5) We compare our results to a semi-analytic galaxy
formation model applied to the Millenium Run simulation.  Differences 
between the data and the model suggest that in the model star formation 
is shut down too efficiently in satellite galaxies.
\end{abstract}

\keywords{galaxies: high-redshift --- galaxies: evolution --- 
cosmology: large-scale structure of the universe}


\section{Introduction}

The clustering of galaxies reflects both the spatial distribution of
dark matter, which is dependent upon cosmological parameters, and the
complex physics which governs the creation and evolution of galaxies
within their host dark matter halos.  Clustering measures can
therefore constrain both cosmology
\citep[e.g.,][]{Peacock01,Abazajian05, Eisenstein05} and galaxy
formation models, including parameterized models like the halo
occupation distribution (HOD) models
\citep[e.g.,][]{Berlind02,Yan03b,Yang05HOD,Zehavi05,Phleps06} or more
physical semi-analytic models and hydrodynamic simulations 
\citep[e.g.,][]{Cen00,Pearce01,Weinberg04}.  In particular, 
the detailed dependence of
clustering on galaxy properties such as luminosity and color at
different redshifts provides a wealth of information on galaxy
evolution.  The formation and evolution of dark matter halos is
straightforward to predict and has been characterized with large
N-body simulations \citep[e.g.,][]{Mo96, Sheth01}, 
but the details of how baryons in galaxies
populate their parent dark matter halos are less well understood and
should depend on the detailed physics of gas
accretion, star formation, and feedback processes.  Comparisons of the
observed dependence of clustering on galaxy properties at different
epochs with theoretical models and simulations can illuminate the 
relative importance of various physical processes involved in galaxy 
evolution.

On scales of $\sim1-20$ \mpch \ the galaxy two-point correlation 
function is roughly a power law,
\xir$=(r/r_0)^{-\gamma}$, and has long been known to depend on galaxy properties,
with one of the strongest dependencies being galaxy type.
The division of the overall galaxy population into two distinct types has
been seen clearly as a function of galaxy morphology, color, and
spectral type, where galaxies tend to be either 
early/bulge-dominated/red/non-star forming or late/disk-dominated/blue/star
forming \citep[e.g.,][]{Strateva01,Madgwick02,Blanton03b,Kauffmann03}.  
Restframe color has been found
to correlate well with both spectral type and morphology, with some
scatter.  In this paper we focus on the
dependence of clustering on restframe color, as this quantity can be 
determined more robustly for large samples of galaxies at both low and 
high redshift than, for example, morphological type.

The correlation between galaxy type and clustering amplitude was first
seen by \cite{Davis76} as a difference in the clustering slope of
morphologically-selected elliptical and spiral galaxies, with
ellipticals having a steeper slope.
This implied that ellipticals are more tightly clustered than spirals
and that a neighboring galaxy of an elliptical was more likely be to
another elliptical.  This was confirmed by \cite{Dressler80} 
who found a tight
relation between morphology and local galaxy density in clusters of
galaxies, with ellipticals preferentially located near the centers of clusters.

The color dependence of galaxy clustering has been measured more precisely
in subsequent surveys of the local universe \citep[e.g.,][]{Loveday95, Hermit96,
Willmer98}, with the most statistically-significant results provided
by the latest large surveys, the 2dF Galaxy Redshift Survey (2dFGRS)
and the Sloan Digital Sky Survey (SDSS)
\citep{Norberg02,Zehavi02,Madgwick032df,Li06clust,Zehavi05}.  Table 1
provides an overview of clustering results for color- and spectral type- 
selected samples from 2dFGRS and SDSS as well as recent intermediate
redshift surveys.

In the 2dFGRS, galaxies are classified into early- and late-type using
a spectral analysis instead of using restframe colors
\citep{Madgwick02}.  \cite{Madgwick032df} find that early-type
galaxies are much more strongly clustered than late-type, with a
relative bias within spheres of 8 \mpch \ radius of $b_{rel}=1.45
\pm0.14$.  \cite{Norberg02} investigate the luminosity dependence of
clustering of early- and late-type galaxies in the 2dF data and find
that while there is a strong difference in the relative mix of early-
and late-types as a function of luminosity, at $L>L^*$ this mix is not
driving the overall increase of clustering for bright galaxies.  Both
galaxy types display a clustering increase at higher luminosities,
with a constant relative bias between early- and late-types as a
function of magnitude at $L>L^*$.

These trends are confirmed by \cite{Budavari03} and \cite{Zehavi04}
using SDSS data. These authors find that both red and blue populations
show a significant increase in the correlation length with luminosity,
though within the blue cloud there is no significant change in the
slope of \xir \ for the brighter galaxies.  \cite{Zehavi04} find that
within the red population, however, the faintest galaxies
($-19<M_r<-18$) have a steeper slope, which is likely due to a higher
fraction of these galaxies being satellites of central galaxies in
groups and clusters \citep{Berlind05}.

Color dependence of galaxy clustering has been detected in a variety of 
intermediate redshift ($0.2<z<1$) surveys
\citep[e.g.,][]{Carlberg97,Carlberg01,Shepard01,Firth02,Phleps06}.  
The most robust results to date at $z\sim1$ are from the VIMOS-VLT
Deep Survey \citep[VVDS;][]{LeFevre05vvds} and the DEEP2 Galaxy
Redshift Survey \citep{Davis03}.  Using VVDS data, drawing from a
total sample of 6500 galaxies at $0<z<1.5$ over $\sim0.5$ deg$^2$,
\cite{Meneux06} measure the clustering of red and blue galaxies,
dividing the two classes using the observed color bimodality.  They
find that red galaxies have a higher correlation length and steeper
slope than blue galaxies at $z=0.8$, and that the relative bias of red
to blue galaxies does not depend significantly on luminosity
\citep{Marinoni05}.
These results are consistent with those of \cite{Coil04xisp}, who use 
an early DEEP2 data sample of 2200 galaxies at $0.7<z<1.35$ in
$\sim0.35$ deg$^2$ to determine that galaxies redward
of the median color of the sample, $(B-R)_0=0.7$, have a significantly 
higher correlation length and steeper slope than bluer galaxies.

Unfortunately, previous studies at intermediate redshift 
covered relatively small volumes and are dominated by large cosmic
variance errors.  Here we improve upon earlier $z\sim1$ DEEP2 results
using the completed DEEP2 data set and volume-limited samples.  We
measure galaxy clustering in several restframe color bins, with a
particular emphasis on `green valley' galaxies at the intersection between 
the red/early- and blue/late-type populations, and as a function of
luminosity within the red and blue galaxy populations separately.

The correlation between galaxy type and clustering strength is also
reflected in measurements of the `environment' or local galaxy overdensity 
around objects of a given class.  Locally,
\cite{Hogg03} find a strong relation between local overdensity and
galaxy color in SDSS data, on scales of both 1 \mpch \ and 8 \mpch, 
with red galaxies favoring more overdense
regions. They find little dependence of overdensity on luminosity
for galaxies in the blue cloud (though the brightest blue galaxies 
do appear to be in more overdense regions) and a strong dependence on 
luminosity for red galaxies locally, with both the brightest and 
faintest red galaxies
being in denser environments than intermediate luminosity red galaxies.  
The lack of a stronger luminosity-overdensity
relation for blue cloud galaxies may be somewhat at odds with the
clustering results from 2dF and SDSS, which find a steeper, roughly linear
trend of relative bias with magnitude, in that brighter
blue galaxies are significantly more clustered than fainter blue galaxies 
\citep{Norberg02,Zehavi05}.  
\cite{Swanson07}, however, also study the clustering of blue galaxies
in SDSS and do not find a significant dependence on the relative bias
with luminosity, in the same magnitude range as these other local studies, 
in contrast to the results of \cite{Norberg02,Zehavi05}.

Galaxy overdensity as a function of color and luminosity was recently
measured for DEEP2 galaxies by \cite{Cooper06}, who find a similar
dependence on color as is seen locally, and a dependence on luminosity
for the brightest of the blue galaxies, which lie in overdense
regions.  No dependence on luminosity was found for galaxies on the
red sequence, in the luminosity range probed by DEEP2.  We expand upon
that study here by measuring the clustering as a function of scale (instead
of using a scale-averaged overdensity) using a statistic that may more
readily be compared with simulations, as it does not depend upon the bias of
the tracer sample used,  as is the case with overdensities.
The two-point correlation function results presented here will be
used for HOD modeling of the galaxy population at $z\sim1$ as a function
of galaxy type in a future paper.  

Significant deviations in the two-point correlation function from a
power-law form have recently been detected on scales $r<10$ \mpch \ for
both local and high redshift populations
\citep[e.g.,][]{Zehavi04,Lee05,Ouchi05,Coil06lum}. These departures from a
power law are naturally explained in the HOD framework, which provides
a statistical understanding of the relation between galaxies and their dark
matter halos, as the transition between pairs of galaxies within the
same halo (the `one-halo' term) on small scales and galaxies in
different halos (the `two-halo' term) on larger scales
\citep{Zehavi04}.  Small-scale deviations from a power law are
expected to be more significant at $z\geq1$, where the exponential 
tail of the mass function moves to smaller masses, so that at a given mass
scale the slope of the mass function becomes steeper \citep{Kravtsov04,Conroy06,Zheng07}.
Observationally, deviations are also seen at a given redshift for the 
brightest galaxy samples, with $L>L^*$.  This is likely due to the 
satellite fraction decreasing for brighter samples; brighter 
galaxies are more likely to be central galaxies than satellite galaxies 
\citep{Zehavi05, Mandelbaum06, vandenbosch07, Zheng07}, which leads 
to a steeper one-halo term (Tinker et al, in prep.) as more pairs 
are central-satellite pairs rather than satellite-satellite pairs and
have, on average, a smaller separation.  

Departures of the two-point correlation function for DEEP2 galaxies from 
a power law as a function of luminosity are presented in
\cite{Coil06lum}; here we perform a similar analysis for galaxies in DEEP2 as
a function of color.  Given that the galaxy correlation function is
not a perfect power law, quoted \rr \ and $\gamma$ fits clearly do not
fully capture all of the information present in clustering measures.
However, they are still useful as an approximate measure of the
amplitude and shape of the correlation function on scales of $\sim1-10$ \mpch \ 
and facilitate simple
comparisons between observational results and models.  As such, 
we provide \rr \ and $\gamma$ fits here but also discuss deviations 
from a power-law form for the data samples investigated.  

The outline of the paper is as follows: \S 2 briefly describes the
DEEP2 survey and defines the data samples used here.  In \S 3 we
discuss the methods used in this paper to measure the two-point
correlation function, infer the real-space correlation length and
slope, and correct for observational biases. \S 4 presents our results
on the color dependence of galaxy clustering at $z\sim1$.  In \S 5 we
compare our results with the semi-analytic models of \cite{Croton06}
applied to the Millenium Run dark matter simulation, and we conclude
in \S 6.


\section{DEEP2 Galaxy Samples}

\subsection{The DEEP2 Galaxy Redshift Survey}

The DEEP2 Galaxy Redshift Survey is a recently-completed project using the
DEIMOS spectrograph \citep{Faber03} on the 10m Keck II telescope to
survey optically-selected galaxies at $z\simeq1$ in a comoving volume of
approximately 5$\times$10$^6$ $h^{-3}$ Mpc$^3$.  Using $\sim1$~hour
exposure times, the survey has measured high-confidence ($>95$\%, 
quality flag $Q=3$ or $4$) 
redshifts for $\sim30,000$ galaxies in the redshift range $0.7<z<1.5$ 
to a limiting magnitude of $R_{\rm AB}=24.1$.  

The survey covers three deg$^2$ of the sky over four widely
separated fields to limit the impact of cosmic variance.  Each field
is composed of 2-4 `pointings', each of which covers an area of
$\sim0.5$ by $\sim0.7$ deg, corresponding to the area covered by the
CFHT12k camera; there are 10 pointings in total.
The DEEP2 spectra have high resolution ($R\sim5,000$), and rms redshift
errors (determined from repeated observations) are $<35$ km
s$^{-1}$.  Details of the DEEP2 observations, catalog construction and
data reduction can be found in \cite{Davis03}, \cite{Coil04}, \cite{Davis05}
and \cite{Davis07}.
K-corrections, absolute $M_B$ magnitudes and restframe $(U-B)$ 
colors have been derived as described in \cite{Willmer06}. We 
do not include luminosity evolution in the K-corrections for $M_B$. 
Absolute magnitudes given in this paper are in the AB system 
and are $M_B-5$ ${\rm log}(h)$ with $h=1$, which for the remainder 
of the paper we simply denote as $M_B$. 

To convert measured redshifts to comoving distances along the line of
sight, we assume a flat \lcdm cosmology with $\Omega_{\rm m}=0.3$ and
$\Omega_{\Lambda}=0.7$.  We define $h \equiv {\rm {\it H}_0/(100 \ km
\ s^{-1} \ Mpc^{-1}})$ and quote correlation lengths, \rr, in comoving
\mpch.

\subsection{Galaxy Sample Definitions}

From the full flux-limited DEEP2 data set we define a variety of 
volume-limited subsamples, each corresponding to some range in 
galaxy $(U-B)$ color and $M_B$ magnitude.
A volume-limited sample is not subject to redshift-dependent selection
effects, 
i.e., a galaxy of given rest-frame properties will be within such a sample
regardless of its redshift.
While volume-limited samples from a given survey will include 
fewer galaxies than flux-limited samples, they are much easier to
interpret and compare with theoretical models, and are therefore used
here.

\begin{figure}
\begin{center}
\scalebox{0.4}{\includegraphics{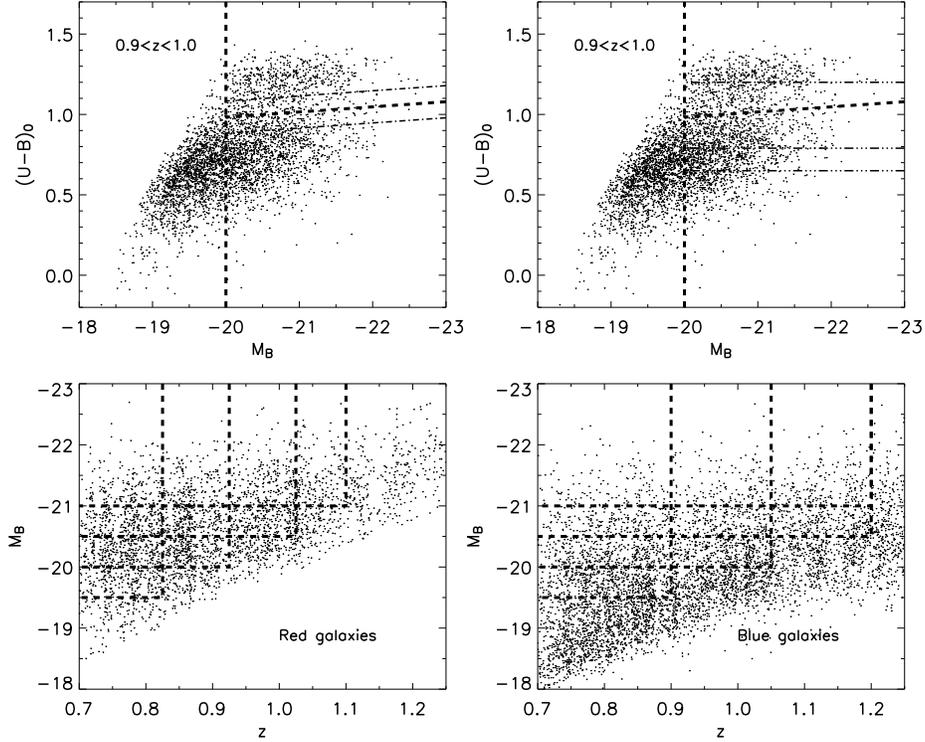}}
\end{center}
\caption{Top row: Restframe color-magnitude diagrams for DEEP2
galaxies in the redshift range $0.7<z<1.0$. The bold vertical dashed
line indicates the $M_B=-20.0$ luminosity cut used for our main 
color samples, while the tilted and horizontal dashed lines show the
various color divisions used to create color subsamples.  The left panel
shows the cuts for the main red and blue samples (bold dashed lines)
and the green sample (thin dashed lines), while the right panel shows
the finer color bins used to further divide the main red and blue
samples.  Bottom row: $M_B$ versus redshift for red(left) and
blue(right) galaxies in the full DEEP2 sample.  The magnitude and
redshift ranges of the red and blue luminosity samples used here are
shown with dashed lines, and properties of each sample are given in
Table 2. The blue galaxy sample has been randomly diluted for clarity 
in this figure. 
\label{mbz}}
\end{figure}

Details of each sample are given in Table 2, and cuts in 
color, magnitude and redshift are shown in Figure \ref{mbz}.  
Following \cite{Willmer06}, we define red and blue
galaxies using the observed color bimodality in 
DEEP2 with the following tilted cut in color-magnitude space 
(in AB magnitudes):
\begin{equation}
(U-B)=-0.032 (M_B+21.62)+1.035.
\end{equation}
We do not allow this color cut to evolve with redshift within the 
redshift range used here.  
The color samples are shown in Figure \ref{mbz}, where the upper
panels are color-magnitude diagrams for all DEEP2 galaxies with
$0.9<z<1.0$.  The {\it main} red and blue samples are
defined as those galaxies lying redward or blueward of the line defined in Eqn. 1,
with $M_B<-20$ (dashed lines in the upper left panel of Figure \ref{mbz})
and in the redshift range $0.7<z<0.925$ for red
galaxies and $0.7<z<1.05$ for blue galaxies.  
While the redshift ranges are not identical for the main red and blue
samples, differences in their clustering properties are due to the
color differences and not the redshift ranges, as we have checked that using 
the same redshift range for both does not change the results.  Using similar
redshift ranges also minimizes the effects of cosmic variance, as any density 
fluctuations will (to first order) be in common between the red and blue samples.
We define green galaxies 
as those within $(U-B)=0.1$ of the red/blue dividing line (shown as dot-dash 
lines in Figure \ref{mbz}) 
in the redshift range $0.7<z<1.0$.
The green galaxy sample contains a subset of galaxies from
each of the red and blue galaxy samples; it is not defined as a distinct 
set of objects, which would require removing green galaxies from the main 
red and blue samples.  

We further divide the red and blue samples into finer color bins: red
galaxies are divided into redder and bluer halves at $(U-B)=1.21$,
while blue galaxies are divided into three color samples at
$(U-B)=0.65$ and $(U-B)=0.79$.  These color cuts are shown in the
upper right panel of Figure \ref{mbz}.  These finer color cuts are 
independent of magnitude and do not have the a tilt in the
color-magnitude diagram, unlike the red/blue dividing line.  Our results do
not change significantly if we instead define all cuts to be dependent on 
both color and magnitude.  The magnitude range we
sample here is not wide, and the number of galaxies near these dividing 
lines is small enough that it does not significantly affect our results.

Within the red and blue galaxy populations we also construct subsamples
as a function of threshold luminosity.  For the brighter luminosity samples
we allow the upper redshift limit to increase with increasing
luminosity, so as to include more galaxies and reduce measurement errors.  
The lower panels of
Figure \ref{mbz} show $M_B$ as a function of redshift for red (left)
and blue (right) galaxies.  The dashed lines indicate the various
luminosity samples defined in Table 2.  However, as illustrated in
Figure 2 of \cite{Gerke07}, the $R_{AB}=24.1$ DEEP2 target selection 
limit defines a restframe color-dependent selection which changes as a
function of redshift and which is not evident in Figure \ref{mbz}.
Figure 2 of \cite{Gerke07} contains $(U-B)$-$M_B$ color-magnitude 
diagrams of DEEP2 galaxies in narrow redshift bins of $\delta z=0.05$.  
The redshift-dependent selection is clearly evident, with the result 
being that red galaxies are not included in the survey at a given $M_B$ 
which is complete for blue galaxies.  This results in a lower upper redshift
limit of the red samples used here compared to the blue samples of
the same luminosity, though the mean redshift of the samples does not vary
by more than $\delta z=0.2$.

Number densities for each sample are given in Table 2; these are derived
from the DEEP2 luminosity function split by color \citep{Willmer06}.
To account for redshift failures we use `minimal' weighting for the 
blue galaxies and `average' weighting for the red galaxies 
(see \cite{Willmer06} for details).  We estimate the number density for 
the green galaxy sample by multiplying the number densities for the main
red and blue samples by the percentage of galaxies in each color that are
in the green valley in the redshift range $0.75<z<0.90$ (12.2\% of 
blue galaxies and 21.6\% of red galaxies).  The dominant error in the
number density estimate derived from the luminosity function is the error
on $\phi^*$; using the errors quoted in \cite{Willmer06}, the fractional
errors on the number densities for the main red and blue samples 
(with $M_B<-20.0$) are $+23\%/-5\%$ for blue galaxies and $+40\%/-55\%$ 
for red galaxies.

\subsection{Coadded Spectra}

\begin{figure}
\begin{center}
\scalebox{0.7}{\includegraphics{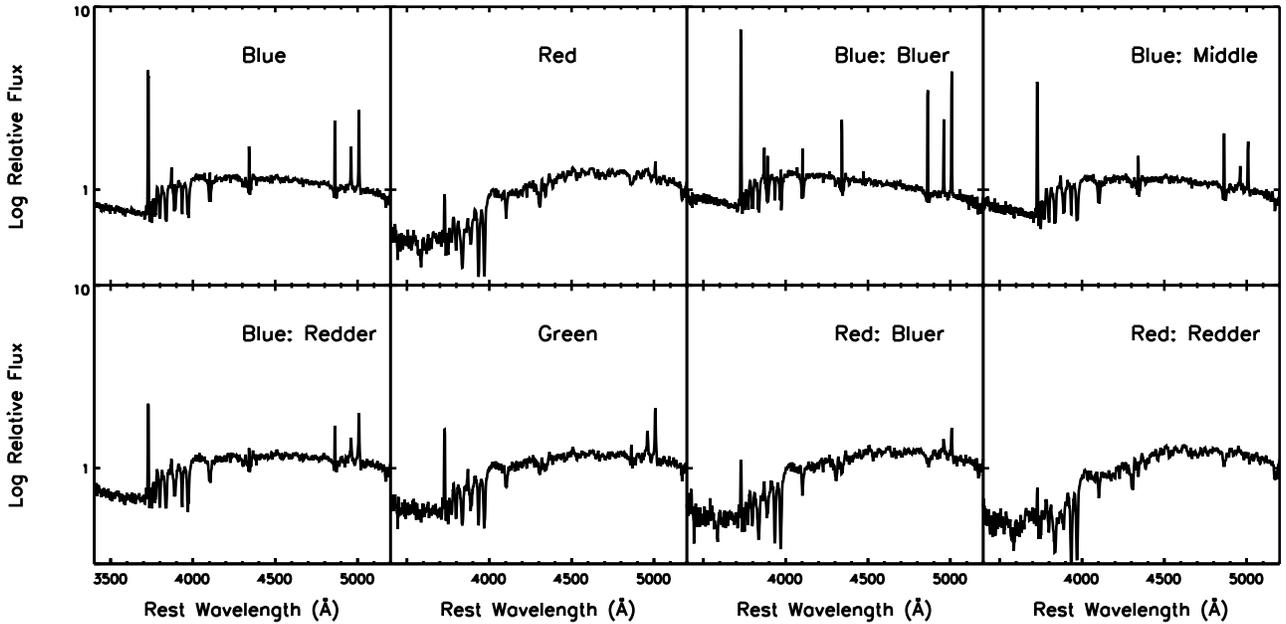}}
\end{center}
\caption{Coadded spectra of galaxies in the various color samples used here.  
The top left panels show average spectra for the main blue and red color samples, 
while the other panels show spectra for the 
finer color bins. The correspondence between restframe color and spectral type is clear.
\label{coadd}}
\end{figure}

Both locally and to at least $z=1$ galaxies are found to be bimodal
not only in terms of their restframe color distribution
\citep[e.g.,][]{Strateva01,Blanton03b,Bell04} but also in 
morphological and spectral type
\citep[e.g.,][]{Madgwick02,Madgwick03deep}. 
Locally, there are strong correlations between morphology, spectral type and
restframe color, with some scatter that likely reflects both intrinsic variation
and the difficulty of unequivocally determining morphological type.  
Defining galaxy samples by spectral type in DEEP2 is not as straightforward
as using restframe colors \citep{Madgwick03deep}, due to the limited spectral 
range of the DEEP2 spectra, which is a consequence of working at high resolution.
We therefore define galaxy samples here by restframe color and luminosity; however,
we show in Figure
\ref{coadd} coadded rest-frame spectra for galaxies in each of our 
color samples in order to investigate and illustrate the physical
nature of the optical emission for galaxies in each sample.
These coadded spectra demonstrate that our color divisions correspond 
directly to classical spectral types. The coadditions correct for the relative
probability that each galaxy was selecteding for observation by DEEP2, but otherwise
all objects are given equal weight.  To maximize signal-to-noise and robustness, 
pixels affected by night sky emission are deweighted and those affected by instrumental
artifacts are removed.

The top row of Figure \ref{coadd} includes the coadded spectra for the main blue and red
galaxy samples.  The blue galaxies display strong emission lines,
primarily from star formation ([OII]~$\lambda 3727$, H$\gamma$,
H$\beta$, [OIII]~$\lambda\lambda 4959, 5007$), along with weak Ca H+K
$\lambda\lambda 3934, 3968$ absorption and stronger Balmer absorption features.
The main red galaxy sample has a small amount of [OII] and [OIII] emission,
primarily from AGN \citep{Yan06}, as well as Ca H+K absorption and
weak Balmer absorption features.  

We also show coadded spectra for galaxy samples selected in finer color
bins.  Within
the blue cloud, the bluer galaxies have much stronger star-forming
emission lines in equivalent width, while
the redder galaxies have lower EW lines; relative to the normalized continuum; 
in luminosity the redder of the blue galaxies have stronger [OII].  

The green galaxies, composed of both 
blue and red galaxies, have lower EW star-forming lines than the blue
galaxies. The green galaxy spectrum does not appear to be a simple mix of the 
red and blue galaxy spectra, as the line ratios are different. 
 The green galaxy spectrum has a much higher [OIII]/H$\beta$
ratio than the blue galaxy spectrum or the average of the `blue:redder' 
and `red:bluer' spectra;  
this indicates a higher fraction of AGN
in the green population relative than in the blue cloud as a whole \citep{Salim07, 
Nandra07}, and also 
shows that the green galaxies are not a simple average of the redder of
the blue cloud galaxies and the bluer of the red sequence galaxies. 
The high [OIII]/H$\beta$ ratio in the green valley population
is likely dominated by the red galaxies in the sample \citep{Weiner07}. 
Within the red sequence, the bluer galaxies have
a higher [OIII]/H$\beta$ ratio, indicating more AGN activity, and the redder galaxies
have stronger Ca H+K absorption features.


\section{Methods}

\subsection{Measuring the Two-Point Correlation Function}

The two-point correlation function \xir \ is defined as a measure of
the excess probability $dP$ (above that for an unclustered distribution) 
of finding a galaxy in a volume
element $dV$ at a separation $r$ from another randomly-chosen galaxy,
\begin{equation}
dP = n [1+\xi(r)] dV,
\end{equation}
where $n$ is the mean number density of the galaxy sample in question
\citep{Peebles80}.

For each galaxy sample we construct a randomly-distributed catalog 
with the same overall sky coverage and redshift distribution as the 
data. We then measure the two-point correlation function using the
\citet{Landy93} estimator,
\begin{equation}
\xi=\frac{1}{RR}\left[DD \left(\frac{n_R}{n_D}
\right)^2-2DR\left(\frac{n_R}{n_D} \right)+RR\right],
\end{equation}
where $DD, DR$, and $RR$ are counts of pairs of galaxies (as a function of
separation) in the data--data,
data--random, and random--random catalogs, and $n_D$ and $n_R$ are the
mean number densities of galaxies in the data and random catalogs.
This estimator has been shown to perform as well as the Hamilton
estimator \citep{Hamilton93} but is preferred here as it is relatively
insensitive to the size of the random catalog and handles edge
corrections well \citep{Kerscher00}.

To estimate the cross-correlation function between two galaxy samples,
we measure the observed number of galaxies from a given sample around
each galaxy in another sample as a function of distance, divided by
the expected number of galaxies for a random distribution.  For this 
we use the simple estimator
\begin{equation}
\xi=\frac{D_1 D_2}{D_1 R_2}-1,
\end{equation}
where $D_1 D_2$ are pairs of galaxies between the two samples and $D_1
R_2$ are galaxy-random pairs between one galaxy sample and the random
catalog of the other sample, where the pair counts have been
normalized by $n_D$ and $n_R$.  

The DEEP2 redshift success rate is $>$70\% overall (defined as the 
percentage of galaxies targeted for spectroscopy that have a well-determined 
redshift) and is not entirely
uniform across the survey; some slitmasks are observed under better
conditions than others and therefore yield a slightly higher
completeness.  We only use regions of the survey with a redshift 
success rate $>$65\%, and the spatially-varying success rate is taken
into account in the window function, which is applied to the 
random catalog to ensure that it has the same spatial distribution as
the survey.  The two-dimensional 
window function of the DEEP2 data in the plane of the sky reflects the 
probability of observing a galaxy and takes into account the overall 
outline of the survey and the geometry of the overlapping slitmasks 
as well as vignetting in the DEIMOS camera and gaps between the DEIMOS CCDs.
The varying redshift success is also taken into account such that 
regions of the sky with a higher
completeness have a correspondingly higher number of random points.
This ensures that there is no bias introduced when computing
correlation statistics.  We also mask the regions of both the random and
mock catalogs where the photometric data have saturated stars or CCD
defects.

Redshift-space distortions due to peculiar velocities along the line
of sight significantly affect 
 estimates of \xir. At
small separations, random motions within a virialized overdensity
cause an elongation along the line of sight (``fingers of god''),
while on large scales, coherent infall of galaxies into potential wells 
causes an apparent contraction of structure along the
line-of-sight \citep{Kaiser87}.
While these distortions can be used to uncover information about the
underlying matter density and thermal motions of the galaxies, they
complicate a measurement of the two-point correlation function in real
space.  In order to
separate the effects of these redshift-space distortions and uncover
the underlying real-space clustering properties, we measure $\xi$ in two
dimensions, both perpendicular to and along the line of sight.
Following \cite{Fisher94}, we define vectors ${\bf v_1}$ and ${\bf v_2}$ to be
the redshift-space positions of a pair of galaxies, ${\bf s}$ to be the
redshift-space separation (${\bf v_1}-{\bf v_2}$), and ${\bf l}
=\frac{1}{2}$(${\bf v_1}+{\bf v_2})$ to be the mean coordinate of  the
pair.  We then define the separation between the two galaxies 
along ($\pi$) and across ($r_p$) the line of sight as
\begin{equation}
\pi=\frac{{\bf s} \cdot {\bf l}}{{\bf |l|}},
\end{equation}
\begin{equation}
r_p=\sqrt{{\bf s} \cdot {\bf s} - \pi^2}.
\end{equation}
To estimate \xisp, we apply the \cite{Landy93} estimator to pair
counts subdivided over a grid in $r_p$ and $\pi$.

\subsection{Deriving the Real-Space Correlations}

While \xisp \ contains useful information about peculiar velocities, 
we also wish to measure the real-space correlation function, \xir.
To recover \xir \ we use a projection of \xisp \  along
the $r_p$ axis.  As redshift-space distortions affect only the
line of sight component of \xisp, integrating over the $\pi$ direction
leads to a statistic \wprp, which is independent of redshift-space
distortions.  Following \cite{Davis83},
\begin{equation}
w_p(r_p)=2 \int_{0}^{\infty} d\pi \ \xi(r_p,\pi)=2 \int_{0}^{\infty}
dy \ \xi[(r_p^2+y^2)^{1/2}],
\label{eqn}
\end{equation}
where $y$ is the real-space separation along the line of sight. 
If \xir \ is modelled as a power law, $\xi(r)=(r/r_0)^{-\gamma}$, then \rr \ 
and $\gamma$ can be readily extracted from the projected correlation
function, \wprp, using an analytic solution to Equation \ref{eqn}:
\begin{equation}
w_p(r_p)=r_p \left(\frac{r_0}{r_p}\right)^\gamma
\frac{\Gamma(\frac{1}{2})\Gamma(\frac{\gamma-1}{2})}{\Gamma(\frac{\gamma}{2})},
\label{powerlawwprp}
\end{equation}
where $\Gamma$ is the usual gamma function.  A power-law fit to \wprp \
will then recover \rr \ and $\gamma$ for the real-space correlation
function, \xir.  

In practice, \xir \ is not expected to be a power law at very large scales 
($>> 10$ \mpch), nor can we measure \xisp \ accurately to 
infinite separations as assumed in Equation \ref{eqn}.
Here we integrate \wprp \ to $\pi_{\rm max}=20$ \mpch, as \xisp \ becomes
noisy at larger separations. 
It is not appropriate to then apply equation \ref{powerlawwprp} directly; 
for power-law correlation functions, that equation becomes increasingly 
inaccurate at large $r_p$  (failing significantly where $r_p/\pi_{\rm max}\gtrsim 0.25$).  
Instead, we must compare the observed \wprp \ to the integral of  \xisp \ over 
separations within $\pi_{\rm max}$, as predicted for a given \rr \ and $\gamma$.

This prediction is complicated by the presence of redshift-space distortions 
due to coherent infall of galaxies
(\cite{Kaiser87}; see \cite{Hamilton92} and section 4.1 of
\cite{Hawkins03} for the relevant equations for correlation
function analyses).  If $\pi_{\rm max}$ were infinite, these distortions would 
have no effect, as they merely change the line-of-sight separations of pairs of 
galaxies; but that is not the case here.  The strength of these distortions 
depends on $\beta=\Omega_m^{0.6}/b$,
where $\Omega_m$ is defined at the mean redshift of the sample (not $z=0$) and
$b$ is the linear bias between the clustering of galaxies and dark matter.

We therefore recover \rr \ and $\gamma$ as follows.  We first measure \wprp \ 
in the data by integrating \xisp \ to $\pi_{\rm max}=20$ \mpch.
We then estimate \wprp \ for dark matter particles at the mean redshift 
of the data sample, using the publicly-available code of \cite{Smith03} 
(not including redshift-space distortions),  
where we integrate the dark matter \xisp \ to $\pi_{\rm max}=20$ \mpch.  
We then estimate the linear galaxy bias from the ratio of these quantities: 
$b^2=[w_p]_{\rm gal}/[w_p]_{\rm dark matter}$.  
We use the average value of this $b$ over scales $r_p=1-10$ \mpch \ and use  
$\Omega_m$ at the mean redshift of the sample (for an LCDM model with 
$\Omega_m(0)=0.3$), to provide an estimate of the redshift-space distortion 
parameter $\beta$.  Using $\Omega_m(0)=0.24$ rather than 0.3 increases our 
recovered \rr \ values by 1\% and does not change $\gamma$.
We can then predict the observed \wprp \ for any point in this grid by integrating 
\xisp \ to $\pi_{\rm max}$=20 \mpch.  The best-fit \rr \ and $\gamma$ values are 
those that minimize the $\chi^2$ difference between the data and the model prediction.

Although our estimated bias, and hence the $\beta$ used in this procedure, may 
not perfectly match the bias corresponding to the best-fit values of  \rr \ and 
$\gamma$, the measured values prove to be insensitive to that estimate.  Increasing 
the bias by 20\% increases \rr \ by 1\% and decreases $\gamma$ by 1\%.
This method for recovering \rr \ and $\gamma$ assumes
that \xir \ is a power law only to a scale of $\pi_{\rm max}$ and results
in \rr \ and $\gamma$ values within a few percent of those obtained using
Equation \ref{powerlawwprp}, for our value of $\pi_{\rm max}=20$ \mpch.
Deviations from Equation \ref{powerlawwprp} are significant only on
larger scales, where $r_p/\pi_{\rm max} \gtrsim 0.25$.
We also consider an alternative method for recovering correlation functions 
in the next section.

Errors on \wprp \ are calculated using the standard error across the
10 separate data pointings.  We do not use a full covariance matrix in 
the fits.  Errors on \rr \ and $\gamma$ are derived
from jacknife resampling of the separate pointings and therefore take into account
the covariance among the $r_p$ bins.

\subsection{Comparison with an Alternative Clustering Statistic}

Given that it becomes necessary to model the effects of the
truncation of $w_p$ and include redshift-space distortions when comparing 
theory with observations, Padmanabhan, White and Eisenstein (2007,
hereafter PWE07) \nocite{nikhil07}
propose an alternative statistic, $\omega(R_{s})$,
which has better convergence properties to the real-space quantities.
This statistic is defined as 
\begin{eqnarray}
\label{eq:omegadef}
\omega(R_s) & \equiv & 2\pi \int r_p\,dr_p\ G(r_p, R_{s}) w_{p}(r_p) \nonumber
\\ & = & 2\pi \int r_p\,dr_p\ G(r_p, R_{s}) \int^{\pi_{max}}_{-\pi_{max}} \,d\pi\
\xi(r_p,\pi) \,\,
\end{eqnarray}
where again $r_p$ and $\pi$ are transverse and line of sight directions, 
respectively. The filter function, $G(r_p, R_{s})$ is chosen to be non-zero 
only for $R_{s} < 1$ and to have zero integral (when integrated against 
an $r^2$ measure).  We follow PWE07 and adopt $G(r_p)=R_s^{-3}
G(x=r_p/R_s)$ with
\begin{eqnarray}
G(x) = & x^{4} (1-x^{2})^{2} \left(\frac{1}{2} - x^{2}\right) & x \le
1 \\ = & 0 & x > 1 \,\,.
\label{eq:gfilter}
\end{eqnarray}
This statistic has the further advantages of being unbinned and
having a well-localized kernel in $\xi(r)$, probing the correlation 
function at a scale $\sim R_{s}$. We note that
$\omega(R_{s})$ can be thought of as performing the Abel inversion
from $w_{p}$ to $\xi$, while avoiding differentiating noisy data.

In order to test the robustness of the methods presented above, below in 
Section 4.1 we
compute $\omega$ for the main red and blue galaxy samples to compare
with the results obtained using the methods in Section 3.2. 
To compute $\omega$ we use the
\cite{Landy93} estimator for the two-dimensional correlation
function, and the techniques in PWE07 to rewrite the integrals in
Eqs.~\ref{eq:omegadef} as weighted pair sums; the line of sight
integral is truncated at $\pi_{max}=40$ \mpch.  
Uncertainties in $\omega$ are estimated by
jackknife resampling the ten DEEP2 pointings. Since $\omega$ is
unbinned, one can choose arbitrarily finely spaced bins, with the
disadvantage that adjacent bins become highly correlated. We have
verified explicitly that our fits for \rr \ and $\gamma$ given below 
are insensitive to the particular choice of binning we adopt.

It is straightforward to relate $\omega$ to the three-dimensional 
correlation function, $\xi$. For a power-law correlation function,
$\xi(r) = (r/r_0)^{-\gamma}$, and our choice of $G(r_{p},R_{s})$, 
this relation becomes 
\begin{equation}
\label{eq:poweromega}
\omega = \frac{2 \pi^{3/2}
\Gamma(\frac{\gamma-1}{2})}{\Gamma(\frac{\gamma}{2})}
\frac{4(\gamma-1)}{(7-\gamma)(9-\gamma)(11-\gamma)(13-\gamma)} \left(
\frac{r_p}{r_{0}}\right)^{-\gamma} \,.
\end{equation}

\subsection{Systematic Biases due to Slitmask Design}

When observing with multi-object slitmasks, the spectra of targets
cannot be allowed to overlap on the CCD array; therefore, objects that
lie near each other in the direction on the sky that is perpendicular to the
wavelength direction on the CCD cannot be simultaneously observed.
This results in undersampling the regions of the sky with the highest
density of targets.  To reduce the impact of this bias, adjacent
DEEP2 slitmasks are positioned approximately a half-mask width apart, giving
each galaxy two chances to appear on a mask; we also adaptively 
tile the slitmask locations to hold constant the number of targets per
mask. In spite of these steps, the probability that a target is
selected for spectroscopy is diminished by $\sim25$\% if the distance
to its second nearest neighbor is less than 10 arcseconds (for
technical details of the slitmask design, see
\cite{Davis03,Davis05}). This introduces a roughly predictable systematic bias
which leads to underestimating the correlation strength on small
scales. The objects that are in conflict are often not at the same
redshift, however, such that the effect is not particularly large.

To correct for this effect we use the mock catalogs of \cite{Yan03}
with galaxy colors added as described in \cite{Gerke07}.  We measure
the ratio of both \xisp \ and \wprp \ in the mock catalogs between 
samples of galaxies from catalogs with and without the slitmask target
selection algorithm applied.  We then multiply the measured \xisp \
and \wprp \ in the data by this ratio, which is a smooth function of
scale and varies from $\sim20$\% on the smallest scales to 2\% on scales 
greater than $r_p=1$ \mpch.  
We tested mock catalog samples with identical redshift, luminosity and
color ranges as each data sample used here and found no significant
differences in the multiplicative corrections as a function of color
or luminosity, such that the same correction derived from the largest
sample is our best estimate for all other samples.
The effect of this slitmask correction is small; it increases \rr \ by
1.5\% and $\gamma$ by 2.5\%.  We include an additional error due to
this correction, which is added in quadrature to the error on
\wprp. This error is estimated from the variance on the correction
in the mock catalogs and is a function of scale, varying from 10\% at
$r_p=0.05$ \mpch \ to 2\% on scales $r_p>1$ \mpch.

\section{Galaxy Clustering Results}

\subsection{Blue and Red Galaxies}

We first compare the main blue and red galaxy samples, both of which
have $M_B<-20$.  The top left panels in Figure
\ref{xisp_all} show \xisp \ for these two samples, with contours
indicating constant probability, where the dark line is $\xi=1$.
There are several clear trends present in
these diagrams; red galaxies are seen to be more clustered than blue
galaxies (a given $\xi$ is seen at larger separations) and show much
stronger fingers of god, as seen in the elongation along the $\pi$
direction at small $r_p$ separations ($<$1 \mpch).  Both of these
results reflect the fact that a color-density relation is in place at $z\sim1$,
with red galaxies residing more in overdense regions such as groups
and clusters than blue galaxies.  Infall on large scales is plainly
seen in these \xisp \ diagrams for blue galaxies, as indicated by the
flattening of the contours along the line of sight. 
Coherent infall may be present
for red galaxies as well, though it is harder to detect due to the
larger fingers of god (see Section 4.9 for more details).
Redshift-space distortions are significant out to $\pi=20$ \mpch, such
that it is important to not use the usual power-law
estimator (Equation \ref{powerlawwprp}) to infer \rr \ and $\gamma$ from \wprp,
as this equation assumes that one has integrated \xisp \ to
$\pi=\infty$.

The top left panel of Figure \ref{wprp_all} shows the projected
correlation function, \wprp, for the main red and blue samples.  Red
galaxies exhibit a significantly steeper slope than blue
galaxies, and both samples are well approximated by a power law within
the error bars.  Power-law fits are given in Table 2, fitting the data
on scales $r_p=0.1-20$ \mpch; the smallest $r_p$ bin (0.05 \mpch) is
not used in these fits due to the rise on small scales which is seen
in some samples (see Section 4.3 for further discussion of deviations
on small scales).  Including the smallest bin in these fits, or
fitting only on scales larger than $r_p=1$ \mpch, does not change the
results for the main blue and red galaxy samples.

The correlation length for red galaxies is found to be significantly
higher than for blue galaxies: $r_0=5.25 \pm0.26$ \mpch \ compared to
$r_0=3.87 \pm0.12$ \mpch.  These values for \rr \ at $z\sim1$ are
somewhat lower than what is found for local samples (Table 1); 
however, \rr \ has only increased by $\sim$5--15\% by $z\sim0.1$ for 
blue galaxies and $\sim$10--20\% for red galaxies.  Furthermore, 
there has been very little evolution in the slope of the
correlation function for either red or blue galaxies since $z\sim1$.
We discuss the implications of this in Section 6.

Our \rr \ values for blue and red galaxies are somewhat higher than those
reported by \cite{Coil04xisp} and \cite{Meneux06}, though generally
within the large errors in those findings.  
\cite{Coil04xisp} use early data from a portion of one DEEP2 field and
\cite{Meneux06} use the VVDS Deep Survey first epoch data; both of these
samples are roughly 1/6 of the dataset used here, and neither 
\cite{Coil04xisp} nor \cite{Meneux06} constructs volume-limited subsamples.  
For blue galaxies, our \rr
\ value found here is 2$\sigma$ and 5$\sigma$ higher than \cite{Coil04xisp} and
\cite{Meneux06}, while for red galaxies our \rr \ value is 1$\sigma$
and 2$\sigma$ higher, respectively, where the errors have been added
in quadrature.  \cite{Brown03} measure the
clustering of red galaxies in the NOAO Deep Wide-Field Survey, using
photometric redshifts.  They find at $z=0.85$ that $r_0=6.7 \pm0.8$
for red galaxies with $-21.5 < M_R < -20.5$ (roughly $L/L*$), 
which is 2$\sigma$ higher than our result.  \cite{Heinis07} measure
the clustering of UV-selected galaxies, again using photometric
redshifts, and find that at $z\sim0.9$ $r_0=4.92 \pm0.5$ \mpch \ and 
$\gamma=1.7 \pm0.09$. Their clustering scale-length is 2$\sigma$ higher 
than what we find here for star-forming galaxies.

We also use the estimator of PWE07, as described in Section 3.3, 
to measure $\omega$ for the main blue and red
samples; the results are shown in Fig.~\ref{omega}.  We have not 
attempted to correct for slitmask effects in the $\omega$ estimator; these effects
are small, as discussed above.  
Using Eqn. \ref{eq:poweromega}, we fit for $r_{0}$ and $\gamma$ for both galaxy samples. 
The resulting power-law fits are plotted in
Fig.~\ref{omega} (solid lines), and are compared with the fits
obtained from fitting to $w_{p}$ (dashed lines). The 
two fits are clearly consistent with each other, given the measurement
errors. For the blue galaxy sample, we obtain $r_{0} = 4.45 \pm
0.54$ \mpch \  and $\gamma = 1.47 \pm 0.11$, while for the red galaxies, we find
$r_{0}=5.94 \pm 0.57$ \mpch \ and $\gamma=1.93\pm0.08$.  
These \rr \ and $\gamma$ values are consistent with the estimates derived above 
from fitting to $w_{p}$.  In fact, because the
correlation coefficient between \rr \ and $\gamma$ is large and negative 
($\sim-0.8$), difference between the results of the two estimators are 
much less significant than the projected errors imply.
The error bars on $r_{0}$ and $\gamma$ are larger using the $\omega$ 
estimator, which is not surprising given that $\omega$ 
is an inversion of $w_{p}$ to a localized integral of $\xi(r)$.

\subsection{Red-Blue Cross-Correlation}

In addition to measuring the auto-correlation of blue and red galaxies
separately, the cross-correlation between the two samples can be used
to further understand the spatial relationship between
blue and red galaxies.  The cross-correlation between two samples
measures the clustering of one type of object around the other and
provides information on the mixing of the populations.  On large, linear
scales, in the two-halo regime, blue and red galaxies should be
well-mixed and trace the same overall large-scale structure, such that
the cross-correlation should equal the geometric mean of the
auto-correlations of each sample.  On small scales, however, within
the one-halo term, deviations from the geometric mean encode
information about the differences in the halos the two samples populate.  
For example, if some halos contain almost
exclusively red galaxies and others only blue galaxies, then the
cross-correlation function would fall below the geometric mean.  The
blue-red galaxy cross-correlation function can therefore be used to
distinguish between different star formation quenching scenarios, as
it traces the distribution of quiescent and actively star-forming 
galaxies around each other.

Here we measure the cross-correlation of the main blue and red galaxy
samples to investigate in particular the mixing of blue and red
galaxies on small, one-halo-dominated, scales ($r_p<1$ \mpch).
The two-dimensional \xisp \ cross-correlation function is shown in the
lower left panel of Figure \ref{xisp_all} and the projected
correlation function, \wprp, is shown in the upper right panel of
Figure \ref{wprp_all}.  The cross-correlation is intermediate between 
the auto-correlation of blue and red galaxies.  The best fit scale length is 
$r_0=4.37 \pm0.41$ \mpch \ and slope is $\gamma=1.83 \pm0.04$.  
Fingers of god are 
clearly seen in \xisp, though the coherent infall on large scales 
appears to be more similar to what is seen for the red sample than 
the blue sample. 

Figure \ref{cross_powerlaw} shows \wprp \ for the main blue and red
galaxy samples and the red-blue cross-correlation function divided by
a reference power law \wprp \ with $r_0=4.0$ \mpch \ and $\gamma=1.8$. 
Also plotted are deviations from this power law for the geometric mean:
\begin{equation}
w_p(r_p)_{red-blue}=\sqrt{w_p(r_p)_{red} w_p(r_p)_{blue}}.
\end{equation}
While the cross-correlation (dot-dash line) is consistent with the 
geometric mean (dotted line) within the error bars, on 
scales $r_p<1$ \mpch \ (the one-halo term) the cross-correlation 
is closer to the blue \wprp \ than the red. This reflects a deficit 
of blue galaxies near red galaxies, which could
be due to the centers of groups being occupied preferentially by red 
galaxies.  This is consistent with the observed cross-correlation 
between DEEP2 groups and galaxies \citep{Coil05}, which is higher for 
red galaxies than blue galaxies, again indicating that red galaxies
tend to be at the centers of groups.

A similar effect has been seen at $z=0.1$ in the 
SDSS red-blue cross-correlation function.
Both \cite{Zehavi05} and \cite{Swanson07} find that the red-blue
cross-correlation amplitude is lower than the geometric mean of
the red and blue clustering on small
scales.  \cite{Wang07} measure the cross-correlation as a function
of luminosity and find a suppression relative to the geometric mean 
for faint red and faint blue galaxies, but not for brighter samples.
\cite{Weinmannconform} find a similar effect in their SDSS group catalog,
which they term `galactic conformity', in which the colors of 
satellite galaxies are correlated with the colors of their central galaxies;
i.e., red satellite galaxies tend to be around red central galaxies, and
the same is seen for blue galaxies.

\subsection{Luminosity Dependence for Blue and Red Galaxies}

The luminosity dependence of clustering in the overall DEEP2 galaxy
sample was presented in \cite{Coil06lum}; here we expand upon that
study by computing the luminosity dependence of clustering within the
blue and red galaxy populations separately.  
The median $M_B$ of each sample is given in Table 2.
For comparison, at $z=0.9$, $M^*$ is 
$M_B=-20.54 (+0.03/-0.01)$ for blue galaxies and $M_B=-20.35 (\pm0.03)$ 
for red galaxies   
\nocite{Willmer06}(from Willmer et al. 2006, where we use the 
`minimal' weighting for blue galaxies and `average' weighting 
for red galaxies).  
The main blue sample has a median $M_B$ value near $M^*$, while the
blue luminosity subsamples range from $0.44$ magnitudes fainter than $M^*$ to 
$0.74$ magnitudes brighter than $M^*$.  The main red sample is 
$0.35$ magnitudes brighter than $M^*$ (in the median) for red galaxies, and the 
red luminosity subsamples range from $\sim M^*$ to $1$ magnitude brighter
than $M^*$.

Figures
\ref{xisp_blue_lum} and \ref{xisp_red_lum} show \xisp \ for four
luminosity threshold samples for blue and red galaxies.  Within the
blue population, there is almost no difference between the faintest
two samples, while the brightest samples may show somewhat stronger fingers 
of god. All of the blue luminosity samples show coherent infall on large
scales.  The red \xisp \ shows almost no difference between any of the
bins; the brightest is noisy, however, due to the small sample size.
Any correlation between the strength of the small-scale 
redshift-space distortion and luminosity within the red galaxy population, 
for the luminosity range that we probe here, is too weak for us to detect. 
\cite{Li06veldisp} find that the pairwise velocity dispersion (see Section
4.8), which quantifies the amplitude of the fingers of god, increases 
with luminosity for blue galaxies in SDSS, while red galaxies do not 
show any luminosity dependence, qualitatively similar to our results at 
$z\sim1$.

The projected correlation function, \wprp, is shown in Figure
\ref{wprp_lum} for blue subsamples (top row) and red subsamples (bottom
row) as a function of luminosity.  Again there is little difference in clustering
between the fainter blue samples (upper left panel); however \wprp \ displays a
significant rise on small scales for the brighter blue samples (upper right panel).  
The red samples show no corresponding rise on small scales for the brighter
samples (lower right panel). Deviations from a power law for the brighter
samples are shown in Figure \ref{wprp_lum_powerlaw}; on scales
$r_p\lesssim0.2$ \mpch \ there is a clear rise in the correlation
function of blue galaxies.  A similar increase on small scales was seen
for all DEEP2 galaxies in \cite{Coil06lum}.
This is likely due to brighter galaxies having, on average, a higher probability of
being a central galaxy rather than a satellite galaxy \citep{Zehavi05, Mandelbaum06, 
vandenbosch07, Zheng07}, which leads to a relative rise in the one-halo term
compared to the two-halo term (Tinker et al, in prep.).  
There may also be a contribution from interaction-driven starbursts \citep[e.g.,][]{Barton00}.
The lack of a strong rise on small scales
in the brighter red samples likely results from red DEEP2 galaxies having a higher
satellite fraction than blue galaxies, at a given luminosity.
We note that while the brighter blue galaxies are more likely to be central galaxies,
they are not necessarily in more massive halos, as shown by the lack of significant
dependence of \rr \ on luminosity within the blue cloud.  

Power-law fits are given in Table 2 where, as above, the scales used to fit 
a power law are $r_p=0.1-20$ \mpch. 
Fitting on scales $r_p=1-20$ \mpch \ instead does not significantly change the results.
The results are shown in the upper panels of Figure \ref{r0_gam_lum}. 
There is no significant
dependence in the correlation length, \rr, or slope, $\gamma$, on 
luminosity for either blue or red galaxies, within the error bars.  

\cite{Cooper06} use local galaxy overdensity
measures to determine the environment of DEEP2 galaxies and conclude  
that within the luminosity range sampled by DEEP2 there is no
significant dependence on luminosity for red galaxies, although there
is for blue galaxies, in that brighter blue galaxies are in more overdense
regions.  Within the luminosity range probed here, the relative bias of
blue galaxies is found by \cite{Cooper06} to increase by $\sim$30\%, which 
is steeper than the trend found here using the correlation function.
The brighter blue samples do show an increase
in the small-scale clustering amplitude relative to the fainter blue samples,
however, which is not reflected in the measured $r_0$ values.  
\cite{Cucciati06} also measure local overdensity at $z\sim0.5-1.5$ in VVDS 
data and find that the fraction of red galaxies at a given overdensity depends 
significantly on luminosity.  We do not find such a trend here with our larger 
dataset.

The lower panels of Figure \ref{r0_gam_lum} compare the clustering scale-length of
red and blue DEEP2 galaxies as a function of $L/L^*$ with local results from 2dF 
\citep{Norberg02} for passive and active galaxies, defined using spectral types, and
from SDSS \citep{Zehavi05}, using restframe colors.  
For this comparison we use $M_B^*=-20.67$ for all galaxies at 
$z=0.9$ \citep{Willmer06} and $M_{b_{j}}^*=-19.66$ and $M_r=-20.44$ 
for all galaxies at $z=0.1$ \citep{NorbergLF,BlantonLF}.  
At $z=0.1$, \cite{Zehavi05,Li06clust,Wang07} all
detect significant luminosity dependence in the clustering of both blue
and red galaxies, while \cite{Swanson07} detect 
luminosity dependence for red galaxies only. \cite{Norberg02} detect
significant luminosity dependence in the clustering of both passive and active
galaxies as well.  

This luminosity dependence for local blue and red galaxies is
significant at $L>L^*$ but not at $L\sim L^*$; there is also a
significant increase in the clustering for faint red galaxies, but
that is below the luminosity range that we probe here.  The DEEP2
sample does not probe $L\gtrsim2 L^*$, such that we can not test the
luminosity dependence at large $L/L^*$.  Within the range that we do
probe, it does not appear that either blue or red galaxies have a
detectable luminosity dependence in their autocorrelation properties.
This is consistent with local results in the same $L/L^*$ range.

Our result that the clustering amplitude does not strongly depend on
luminosity for red galaxies at $z\sim1$ is at odds with
\cite{Brown03}, who find a significant luminosity dependence in the
clustering of red galaxies at $z=0.3-0.9$, for samples between
$M_R=-20$ and $M_R=-22$ (in Vega magnitudes), using photometric
redshift data.  $M^*_R$ for the \cite{Brown03} sample is $\sim-21.0$
(M. Brown, private communication).  Their value of $r_0$ at $z\sim0.6$
for $L/L^*\sim1$ red galaxies is $6.3 \pm0.5$ \mpch, which is only
somewhat higher than that found here ($r_0=5.25 \pm0.26$ at
$\overline{z}=0.82$) or at $z\sim0.1$ by \cite{Zehavi05} ($r_0=5.67
\pm0.37$).  However, they find a steep trend in $r_0$ with $M_R$, with
$r_0$ ranging from $4.4 \pm0.4$ \mpch \ to $11.2 \pm1.0$ \mpch \ for
$L/L^*\sim0.4-2.5$, which is not found here or locally.  No value of
$M^*_R$ would lead to a consistent trend of $r_0$ with luminosity when
compared with our results or local results from \cite{Zehavi05} or
\cite{Norberg02}.  Estimating \rr \ from an angular correlation
function, as done in \cite{Brown03}, requires very accurate knowledge
of the redshift distribution; this may account for the discrepancy
with the results from spectroscopic surveys.

\cite{Coil06lum} find that within the full DEEP2 galaxy sample the clustering
strength depends strongly on luminosity.  
This can still occur even if there is no luminosity dependence for red or blue
samples separately, as the red galaxy fraction is a function of luminosity; 
the higher prevalence of red galaxies in the
brighter samples leads to \rr \ increasing as $M_B$ decreases.

\

\subsection{Green Galaxies}

The clear bimodality both locally and at $z=1$ in galaxy properties, 
whether defined by restframe color, morphology or
spectral type, raises the question of how these two main types arise
and what, if any, evolutionary connection exists between them.  The
observed buildup of galaxies on the red sequence since $z=1$
\citep[e.g.,][]{Bell04,Faber06,Brown07} suggests that blue galaxies
are moving to the red sequence with time, though the details of this
transition are not clear.  The general outline, however, is that star
formation begins to shut down or be quenched in blue galaxies, after which
they passively evolve onto the red sequence.  The star formation
quenching process(es) involved may be causally related to the environment 
of the galaxy (e.g., ram pressure or tidal stripping of gas \citep{Gunn72,Byrd90}), 
such that galaxies in higher density regions end their
star forming phase more readily, thus creating a color-density relation.
Alternatively, the process may be inherent to the galaxy and depend on its age or
stellar or halo mass, in which case the process may be correlated with 
environment but not caused by it \citep{Bundy06, Cooper07}.  
To elucidate what is causing this change, we can 
investigate the clustering properties of galaxies located at the
transition region between these populations.  Here we study `green'
galaxies located near the minimum of the observed color bimodality as defined
in \S 2.2.

\begin{figure}
\begin{center}
\scalebox{0.35}{\includegraphics{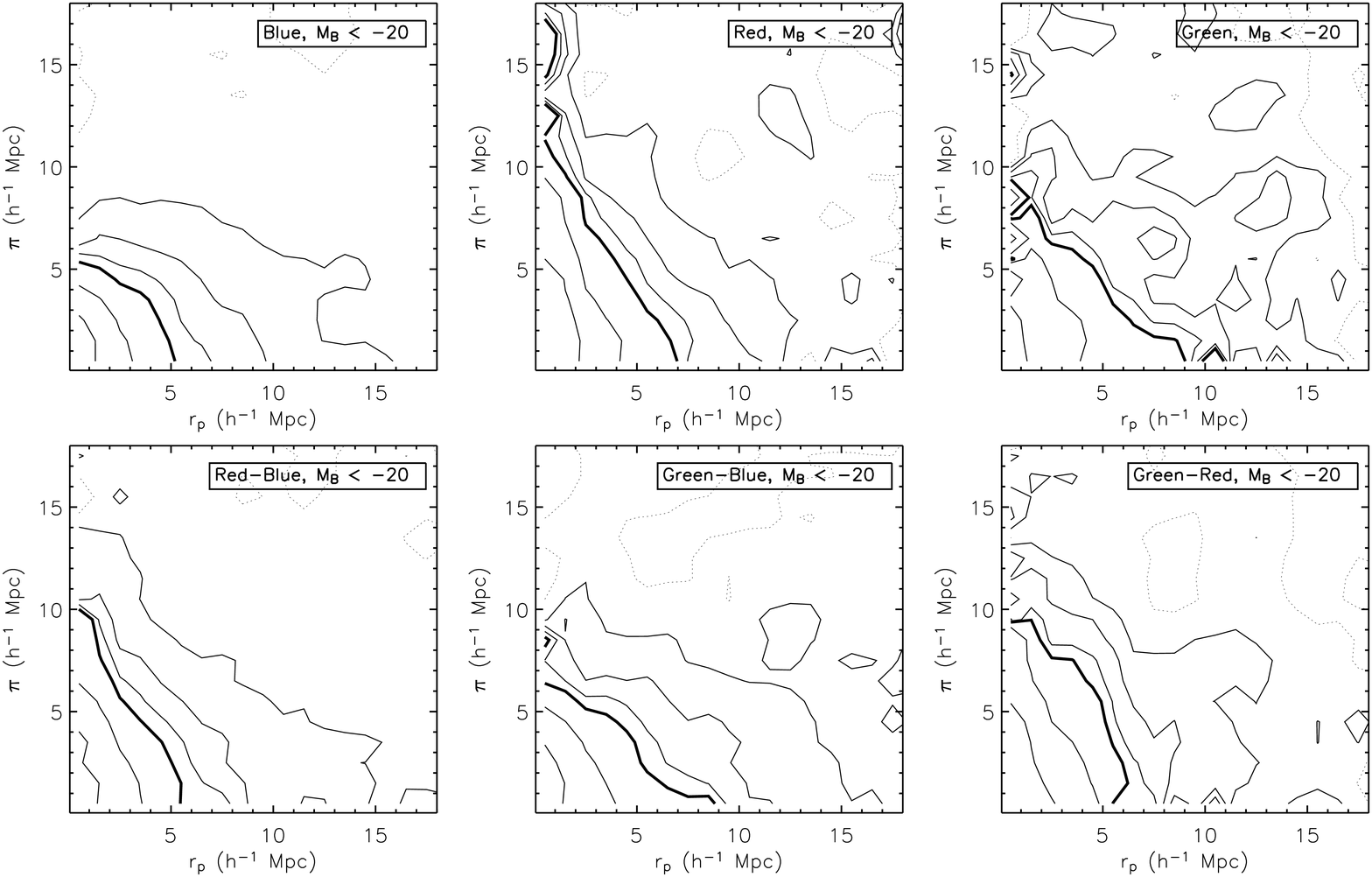}}
\end{center}
\caption{Contours of constant correlation strength 
for the two-dimensional correlation function, \xisp,
for the main blue, red and green samples (top row) and various cross-correlation functions (bottom
row).  A $1\times1$ \mpch \ boxcar smoothing has been applied in the figure, though it is not used in
calculations.  Contours levels are 0.0 (dashed), 0.25, 0.5, 0.75, 1.0 (bold), 2.0 and 5.0.
\label{xisp_all}}
\end{figure}

The green galaxy \xisp, shown in the upper right panel of
Figure \ref{xisp_all}, exhibits an intermediate clustering amplitude
between the blue and red galaxy samples, as well as intermediate-strength 
fingers of god.  The green population displays 
somewhat similar infall on large scales as blue galaxies, though
the contours are noisy on large scales due to the relatively small sample 
size.

The projected correlation function for green galaxies is shown in the lower 
left panel of Figure \ref{wprp_all}, plotted with \wprp \ for red and blue
galaxies for comparison.  On large scales ($r_p>1$ \mpch) the green 
galaxies have a clustering amplitude similar to that of red galaxies, while 
on small scales the amplitude is closer to that of blue galaxies.  
The large-scale agreement with the red galaxy population implies that 
green galaxies reside in or on the outskirts of halos of similar mass as 
red galaxies, as they have the same clustering for the two-halo term.
Green galaxies are therefore generally in the same overdense regions as 
red galaxies.  However, the one-halo term for green galaxies is lower than
for red galaxies, which likely reflects a lower radial concentration of green
galaxies within their parent halos. 
The overall picture is consistent with galaxies having green colors while on the 
outskirts of halos and becoming redder as they reach the center of the 
overdensity.  These results point towards green galaxies being a distinct population
and not a simple mix of red and blue galaxies.

To investigate these trends further, we compute the cross-correlation
between the green galaxy sample and the main blue and red galaxy
samples.  The \xisp \ diagrams (lower right panels of Figure \ref{xisp_all}) 
show more significant infall on large scales in the green-blue cross-correlation
than the green-red cross-correlation (comparing, for example, the $\xi=1$
contours).  The green-red cross-correlation
is very similar to the blue-red cross-correlation (though with smaller
fingers of god), indicating that green galaxies have comparable kinematics 
to blue galaxies.  
There are also smaller fingers of god in the green-red cross-correlation 
than in the red auto-correlation.  
These results are consistent with green galaxies residing 
in the same overdense regions as red galaxies, on average, but with a
lower probability of being at the cores of 
those overdensities, compared to red galaxies.  
This would result in green galaxies having smaller fingers of god than
red galaxies, and a lower auto-correlation on small scales.  It is possible
that green galaxies reside on the outskirts of the same halos that red galaxies
occupy but are still falling in towards the centers of the halos.

The bottom right panel of Figure \ref{wprp_all} shows the projected cross-correlation 
functions between green and blue or red galaxies. The green-red galaxy clustering
amplitude is higher than the green-blue galaxy amplitude on all scales. On scales
$r_p=1-10$ \mpch \ the green-red sample is $39 \pm5$\% more clustered. 
On small scales, dominated by 
the one-halo term, the green-red amplitude is a few times higher than the
green-blue amplitude: a factor of $2 \pm0.2$ times higher on scales 
$\sim0.5$ \mpch \ and a factor of $2.8 \pm0.4$ times higher on scales
$\sim0.25$ \mpch.  One would therefore be {\it much} more likely to find a red
close neighbor (or satellite) galaxy to a green galaxy than a blue neighbor, 
and vice versa, if red and blue galaxy samples were of equal size.  
Another way to frame these results is that a green galaxy is less 
likely to be a central galaxy than a red galaxy of 
similar luminosity, but more likely than a comparable blue galaxy.

We note that the green galaxy \wprp \ likely can not be fit with an 
HOD model that assumes that the radial distribution of galaxies follows that
of dark matter particles or subhalos.  The green galaxy radial distribution is clearly 
different from that of either red or blue galaxies, and most HOD models
do not allow the radial distribution to vary for different galaxy types.

\subsection{Finer Color Bins}

We next divide the blue and red galaxy samples into finer color bins
to investigate the strength of clustering as a function of color
{\it within} each population.  Figure \ref{r0_gam_color}
shows \rr \ and $\gamma$ as a function of color for both the main blue
and red samples (triangles), the green sample, and for finer color bins.
As explained in Section 2.2, the finer color bins are not all
independent; this is shown in the figure with a dotted horizontal line
showing the color range for each point (for clarity the color ranges
for the main blue and red samples are omitted).
In particular, the green galaxy sample contains subsamples of both
blue and red galaxies.

Within the red population (where we have two independent samples)
neither \rr \ or $\gamma$ depend significantly on color; across
the red sequence we detect no change in the clustering
properties of galaxies.  Within the blue population, however, there is
a strong dependence; redder galaxies are more
clustered than the bluer galaxies.  There is no corresponding trend
seen in the slope of the correlation function.
The abrupt change in slope between blue and red galaxies
likely reflects the different radial concentrations and satellite
fractions of the two populations and may indicate that star formation quenching
is an effect associated with the halo properties of galaxies.

Interestingly, the green galaxy \rr \ is found to be the same as the
red galaxy \rr \ (within 1 $\sigma$), while the green galaxy $\gamma$ 
is the same as for blue galaxies, not red galaxies.  This is consistent 
with the interpretation in the previous section that green galaxies 
reside in similar mass halos as red galaxies, and are found generally in
overdense regions, but are less concentrated within their host halos than red
galaxies.  They apparently do not populate the centers of
overdensities.

As discussed earlier, luminosity function studies find that a
significant fraction of the progenitors of local galaxies on the red 
sequence must have been in the
blue cloud at $z\sim1$ \citep{Bell04,Brown07,Faber06}.  Our clustering
results imply that the galaxies that migrate to the red sequence from
$z=1$ to $z=0$ are amongst the redder of the blue galaxies, and the green
galaxies at $z=1$, which are already as clustered as galaxies on the
red sequence.

These trends of finding a color-density relation within the blue cloud
but not within the red sequence are similar to what has been locally
using galaxy environment measures \citep[e.g.,][]{Hogg03,Hogg04}.
There are no other published $z\sim1$ clustering results with such fine
color bins to compare to, and we defer a comparison with the DEEP2
environment results of \cite{Cooper06} to Section 4.7.

\subsection{Galaxy Bias and Relative Bias Between Samples}

To facilitate comparisons to published clustering results measured at other
redshifts or with different statistics than the two-point correlation
function, we calculate the relative bias between various galaxy color
samples.  We define the relative bias as the square root of the ratio
of \wprp \ for two samples.  As the relative bias is scale-dependent,
we calculate the mean relative bias over two scales, $r_p=0.1-15$ \mpch
\ ('all scales') and $r_p=1-15$ \mpch \ ('large scales'), which have 
mean scales of $r_p=4.4$ \mpch \ and $r_p=6.5$ \mpch, respectively.

The relative bias of red to blue galaxies is $b_{rel}=1.28 \pm0.09$ on
large scales and $b_{rel}=1.44 \pm0.07$ on all scales, reflecting the
rise in the relative bias on small scales, $r_p<1$ \mpch. We do not find
any dependence on luminosity for the relative bias between red and blue 
galaxies.  This is in contrast to the results of \cite{Cucciati06}, who
measure the color-density relation at $z\sim0.5-1.5$ in VVDS data and 
find that the color-density relation steepens for brighter galaxies.  No
such trend is detected here.  

The lack of luminosity dependence in the relative bias found here 
could indicate that the ratio of the satellite fraction in red versus
blue galaxy samples does not depend on luminosity.  \cite{Berlind05} show 
that for central galaxies there is no color-density relation; this 
relation only exists for satellite galaxies.  This implies that the
relative bias of red to blue galaxies should be a function of luminosity
if the central to satellite galaxy fraction changes with luminosity 
differentially for red and blue galaxies.  The fact that the data show
no dependence in the relative bias on luminosity could imply that the 
ratio of the satellite fraction in red versus blue galaxy samples is fixed and 
does not depend on luminosity.

The relative bias of red to green galaxies is
$b_{rel}=1.00 \pm010$ on large scales and $b_{rel}=1.12 \pm0.11$ on
all scales. Given that red and green galaxies have different correlation slopes
and redshift-space distortions, the relative bias does not fully
reflect the differences between these populations. It does, however,
indicate that when two-halo-dominated scales are included, on average the
clustering amplitude of green galaxies is perfectly consistent with 
that of red galaxies and is not lower.  The relative bias of green to blue
galaxies is found to be $b_{rel}=1.30 \pm0.11$ on both large and all
scales, with no scale-dependence.

The relative bias found here between red and blue galaxies at $z\sim1$ is
very similar to what is measured at lower redshift (see Table 1; 
$b_{rel}$ is $\sim0-10$ \% lower at $z\sim1$).
\cite{Madgwick032df} compare the clustering of early and late type galaxies 
(defined by spectral type) and find a relative bias of $b_{rel}=1.45 \pm0.14$
integrated to scales of $r=8$ \mpch, while \cite{Willmer98} find a
relative bias of $\sim1.4$ between red and blue galaxies. 
\cite{Zehavi02} find using
SDSS data that the relative bias of red to blue galaxies is $\sim1.6$,
but is highly scale-dependent.  This agreement with lower redshifts
implies that the dependence of color on density at  $z=1$ is as steep as it is
locally; the physical mechanisms responsible for the
color-density relation are just as effective before $z=1$ as they are
after.  The color-density relation was therefore established before $z=1$ 
and subsequent changes may be absolute (in terms of the zeropoint of 
the relation) but not relative (which would change the slope), within
the errors.  As discussed in \cite{Cooper06}, given the low 
abundance of rich clusters at $z=1$, the color-density relation seen 
in the DEEP2 data can not be the result of cluster-specific physics 
such as ram pressure stripping of gas or galaxy 
harassment \citep[e.g.,][]{Gunn72,Byrd90}.

For a given cosmology we can also compute the {\it absolute} galaxy
bias relative to the underlying dark matter density field using
results from N-body simulations.  We use the power spectrum of
\cite{Smith03} to estimate the dark matter clustering amplitude at the
same redshift as the data, for a cosmology with $\Omega_m=0.3,
\Omega_\Lambda=0.7, \sigma_8=0.9$, and $\Gamma=0.21$.  This code is
an analytic fit motivated by the halo model; we include a 5\% error on
the dark matter \wprp \ at each scale in our calculation of the 
absolute bias to reflect the uncertainty in the fit.  
The resulting absolute bias values
averaged on scales $r_p=1-10$ \mpch \ are listed for each color and magnitude
sample in Table 2. For $\sigma_8=0.8$ the
large-scale bias is 13\% higher, while for $\sigma_8=1.0$ it is 10\%
lower.  Note that the absolute bias and
relative bias are measured on different scales.  

We find that the absolute bias is $b=1.65 \pm0.15$ for red galaxies and
$b=1.28 \pm0.04$ for blue galaxies, both with $M_B<-20.0$.  This is
consistent with \cite{Marinoni05}, who find that red galaxies with $M_B<-20.0$ at
$z\sim0.8$ have an absolute bias of $b=1.5 \pm0.6$ and blue galaxies
have $b=1.1 \pm0.6$, measured on scales of $r=8$ \mpch.
The absolute bias of galaxies at $z\sim1$ measured here is higher than at $z=0$
\citep[e.g.,][]{Verde01,Hoekstra02,Seljak05,Simon07}, where galaxies
near $L^*$ are found to have a bias of $b\lesssim1.0$.

\

\subsection{Comparison to Environment Studies}

To compare our results on the color dependence of galaxy clustering
with the DEEP2 environment results of \cite{Cooper06}, 
we show in Figure \ref{relbias} the relative bias
of each color sample used here, relative to the main blue sample,
measured on scales $r_p=1-5$ \mpch.  The main red and blue samples are shown
as triangles, while the finer color bins are shown as diamonds. 
Crosses with dotted error bars are the relative galaxy overdensity as 
measured by the
environment statistic $\delta_3$ (described in \cite{Cooper06}) for
galaxies with $z=0.75-1.0$ and $M_B<-20.0$, as a function of color,
again normalized to the main blue sample. Note that this redshift and
magnitude range are not exactly the same as those used in \cite{Cooper06},
therefore our Figure \ref{relbias} does not look identical to Figure 5
in \cite{Cooper06}. 

The overall agreement is very good between the relative overdensity
and the relative bias from the clustering measures presented here;
these two statistics should be similar but not identical.  The
overdensity estimate,
$\delta_3$, is not measured at a fixed scale but is generally in the
range $r=1-3$ \mpch; we compare to the relative bias from clustering 
averaged on scales $r_p=1-5$ \mpch \ to minimize 
systematics from scale dependence.  The environment
overdensity statistic is in some ways a scale-averaged clustering
measure, so the agreement between the two should be good.  However,
the environment statistic uses {\it all} neighboring galaxies to
define the local overdensity, not just those with the same color or
magnitude as the galaxy for which one is measuring the environment.
This is different from the clustering statistic we use, where only
galaxies within a given color and magnitude range are used for the
auto-correlation function.  Using all galaxies as a density tracer 
could potentially cause the environment statistic to be more 
sensitive to group membership and corresponding small-scale overdensities 
for populations which tend to be central galaxies instead of
satellite galaxies (i.e., if there is only one galaxy of that type 
in an overdensity).

As seen in Figure \ref{relbias}, the agreement between the two
statistics is good but not exact.  
In particular, the clustering results show a continued trend within
the blue population of the bluest galaxies being the least clustered, which
is not clearly reflected in the environment measure, though the discrepancy 
is at the 1$\sigma$ level; the broader environment samples in \cite{Cooper06}
 do show such a downturn.

Our clustering-based results on the color-density relation at $z\sim1$ are very similar to 
local environment findings in SDSS.  \cite{Hogg03,Hogg04} find that SDSS 
galaxies also show a color-density relation within the blue population, 
but not within the red sequence.  They also find that locally the 
overdensity of blue galaxies shows little luminosity dependence, while 
there is strong luminosity dependence within the red galaxy population. 
Unfortunately, the DEEP2 volume is not large enough to provide a fair sample 
of the brightest, rarest red galaxies at $z\sim1$, and the error bars on our
results for $r_0$ as a function of luminosity within the red sequence are 
5-15\%, which may be large enough to mask a modest luminosity dependence for the
range of luminosities we probe.  The general conclusion is that there has not 
been qualitative evolution in the color-luminosity-overdensity space since $z\sim1$.

\subsection{Pairwise Velocity Dispersion}  

On small scales ($r\lesssim1$ \mpch) random motions within galaxy
groups and clusters lead to elongations of the \xisp \ contours along
the line of sight, the so-called ``fingers of god''.  These motions can 
only be seen in spectroscopic redshift surveys, and at $z\sim1$ only
the DEEP2 and VVDS surveys can currently detect these distortions in \xisp.
The virialized motion 
of the galaxies can be measured using the pairwise velocity dispersion,
$\sigma_{12}$, which is estimated by modeling \xisp \ as a
convolution of the real-space correlation function, \xir, with a
broadening velocity function.  The relation between \xisp \ and
$\xi(r)$ is usually assumed to be
\begin{equation}
1+\xi(r_p,\pi)=\int f(v_{12})\left[1+\xi(\sqrt{r_p^2+y(v_{12},\pi)^2})
\right]dv_{12},
\label{sigmamodel}
\end{equation} 
where $f(v_{12})$ is the distribution function of the relative
velocity difference between galaxy pairs along the line of sight 
\citep[e.g.,][]{Davis83}.
This velocity difference is defined as
\begin{equation}
v_{12} \equiv \pi - y+{\overline v_{12}}(r)
\end{equation}
where ${\overline v_{12}}(r)$ is the mean radial pairwise velocity of
galaxies at separation $r$ and $y$ is the real-space separation along 
the line of sight. Here $y$ has units of \kms \ instead of
\mpch \ (this conversion includes a factor of $H_0$, with $h=1$, 
and a factor of $(1+z)$), such that the same physical velocity at $z=1$
has less of an effect on \xisp \ than at $z=0.1$.

To measure the small-scale velocity dispersion on scales $r\sim1$ \mpch, 
we follow \citet{Fisher94} and in Equation \ref{sigmamodel} we use
\begin{equation}
\xi(r_p=1,\pi) \equiv 0.5[\xi(r_p=0.5 \mathrm{Mpc}\,
h^{-1},\pi)+
\xi(r_p=1.5 \mathrm{Mpc} \,h^{-1},\pi)]
\end{equation}
in both real and redshift space, for values of $\pi \le$ 20 $h^{-1}$
Mpc.  We then normalize $\xi(r_p=1,\pi)$ so that our subsequent
fitting will be sensitive to the overall shape of $\xi(r_p=1,\pi)$ but
insensitive to the amplitude.  

An exponential form is usually adopted for $f(v_{12})$
\citep[e.g.,][]{Davis83,Fisher94,Diaferio96,Sheth96}:
\begin{equation}
f(v_{12})={1\over \sqrt{2}\sigma_{12}} \exp \left(-{\sqrt{2}\over
\sigma_{12}} \left| v_{12}-\overline{v_{12}}\right| \right),
\end{equation}
where $\overline {v_{12}}$ is the mean and $\sigma_{12}$ is the
dispersion of the pairwise peculiar velocities.  We assume an infall
model based on the similarity solution of the pair conservation
equation \citep{Davis77},
\begin{equation}
\overline{v_{12}}({\bf r})= -{y\over 1+(r/r_{\star})^2},
\end{equation} 
where $r_{\star}=5$ \mpch \ and $y$ is the radial separation in the
real space.  The results are relatively insensitive to the assumed
model; $\sigma_{12}$ is $\sim$20-40\% lower if $\overline{v_{12}}$ is
assumed to be negligible.

To estimate $\sigma_{12}$, \xir \ is modelled using the power-law fits
for \rr \ and $\gamma$ for each sample, and we minimize $\chi^2$
between the observed $\xi(r_p=1,\pi)$ and modelled $\xi(r_p=1,\pi)$
for a range of $\sigma_{12}$ values.  The results are given Table 2,
where errors are derived from jacknife samples.  Red galaxies are
found to have $\sigma_{12}=530 \pm50$ \kms, while blue galaxies have a
significantly lower value of $\sigma_{12}=240 \pm20$ \kms.  
The error bars are too large to
detect significant differences as a function of luminosity within the
red or blue galaxy populations or as a function of color for the finer color
bins.  Green galaxies have a large value, $\sigma_{12}=490 \pm110$
\kms, consistent with red galaxies and 2$\sigma$ larger than
blue galaxies.  

These numbers are similar to, but somewhat smaller
than, values found at $z=0.1$.  Using the same infall model,
\cite{Zehavi02} measure $\sigma_{12}\sim650-750$ \kms \ for red
galaxies in SDSS and $\sigma_{12}\sim300-450$ \kms \ for blue
galaxies at $z=0.1$. \cite{Li06veldisp} also find for SDSS galaxies that
 $\sigma_{12}\sim600-800$ \kms \ for red galaxies and 
$\sigma_{12}\sim200-400$ \kms \ for blue galaxies.
\cite{Madgwick032df} measure similar values for 
galaxies in the 2dF Redshift survey: $\sigma_{12}=612 \pm92$ \kms \ 
for passive galaxies (defined by spectral type) and $\sigma_{12}=416 \pm76$ 
\kms \ for active, star-forming galaxies.

Our finding that red galaxies have stronger fingers of god at $z\sim1$ 
than blue galaxies again reflects the result that, on average, red
galaxies reside in more overdense regions, such as galaxy groups, than 
blue galaxies.

\subsection{Multipole Moments}  

Peculiar velocities also affect \xisp \ on large scales, where
coherent infall of galaxies onto forming structures flattens \xisp.
A standard method for
quantifying these large-scale redshift-space distortions is to measure
the ratio of the quadrupole to monopole moments of the two-point
correlation function \citep{Hamilton92}.  The two-dimensional 
correlation function can be decomposed into a sum of Legendre polynomials,
\begin{equation}
\xi(r_p,\pi)=\sum_l \xi_l(s){\cal P}_l(\mu),
\end{equation}
where ${\cal P}_l$ is the $l^{th}$ Legendre polynomial and $\mu$ is
the cosine of the angle between the line of sight and the redshift
separation vector, ${\bf s}$.
The multipole moments are defined as
\begin{equation}
\xi_l(s)=\frac{2l+1}{2} \int_{-1}^{1} \xi(r_p,\pi) {\cal P}_l(\mu)d\mu.
\end{equation}
Following \cite{Hamilton92}, we define
\begin{equation}
Q(s) \equiv \frac{\xi_2(s)}{(3/s^2)\int_{0}^{s} 
\xi_0(s')s'^2ds' -\xi_0(s)}.
\end{equation}

This ratio depends on 
$\beta\equiv\Omega_M^{0.6}/b$, where $b$ is
the linear galaxy bias \citep{Kaiser87}, as
\begin{equation}
Q(s)=-\frac{(4/3)\beta+
(4/7)\beta^2}{1+(2/3)\beta+(1/5)\beta^2}.
\end{equation}

Figure \ref{qratio} shows $Q(s)$ for the main blue and red galaxy
samples.  $Q(s)$ is positive on small scales where the fingers of
god are strong and negative on large scales where coherent infall
dominates.  The dotted lines indicate $Q$ predicted on linear scales 
(from Equation 22), assuming 
$\Omega_m$ at the mean redshift of each sample (for the blue sample 
$\Omega_m(z=0.90)=0.75$ and for the red sample $\Omega_m(z=0.82)=0.72$, 
corresponding to 
$\Omega_m(z=0)=0.3$) and using the inferred linear galaxy bias for each
sample, $b=1.28$ for blue galaxies and $b=1.65$ for red galaxies. 
Using $\Omega_m(z=0)=0.24$ increases the predicted $Q$ values by
$\sim$5\%.  For the blue sample $\beta=0.66$ and for the red sample $\beta=0.50$.
Both galaxy samples have negative $Q$ values, on scales
$s>5$ \mpch \ for blue galaxies and $s>10$ \mpch \ for red galaxies,
indicating coherent infall, though the significance is lower for red 
galaxies.  On the largest scales $Q$ increases; however, the error bars
increase on these scales as well, which approach the transverse scale of the
survey and so may be susceptible to systematic effects.  
On scales $s\sim10-15$ \mpch \ the blue
sample is consistent with the predicted constant $Q$ value within the
noise.  The red sample does not reach the predicted negative value 
on the largest scales; however, given the non-negligible pairwise velocity 
dispersion we do
not expect this model to be perfect, as even on scales of 15 \mpch, 
$Q(s)$ is sensitive to $\sigma_{12}$ and the form of $f(v_{12})$.  
This is especially true for the
red galaxy sample, where $\sigma_{12}$ is greater.  
The fact that infall is seen indicates that the clustering 
of galaxies is dynamic and growing, as expected.

\section{Comparison with Semi-Analytic Millenium Run Simulation}

We compare our clustering results for blue and red galaxies with the
recent semi-analytic galaxy formation model of \cite{Croton06} applied
to the Millennium Run N-body dark matter simulation
\citep{Springel05b}.  The Millennium Run follows the dynamical
evolution of $10^{10}$ dark matter particles in a periodic box of
side-length $500\,h^{-1}$Mpc with a mass resolution per particle of
$8.6\times 10^8\,h^{-1}{\rm M}_{\odot}$.  Cosmological parameters
consistent with those measured from the first year WMAP data
\citep{Spergel03, Seljak05} were used.  The galaxy formation model
follows the growth of approximately 25 million galaxies from their 
birth to
the present day and has been tuned to provide a good match to many
observed properties of local galaxies, primarily the luminosity
function and Tully-Fisher relation.  Within this model the behavior
of the observed global galaxy two-point correlation function is well
reproduced at $z=0$, however when split by galaxy color, $z=0$ red galaxies
are somewhat too clustered in the model while blue galaxies are
under-clustered \citep{Springel05b}. Ours is the first test of the
clustering of Millennium Run galaxies as a function of color at 
higher redshift.

Using an ensemble of 15 light-cones, each of which matches the geometry
of one of the four DEEP2 survey fields 
\citep[see][for further details]{Manfred06} 
we measure the clustering of blue and red galaxies with the same $M_B$
and $z$ ranges as our main DEEP2 color samples.  We define galaxies to
be blue or red in the model using the same color cut as used for DEEP2
galaxies (Equation 1).  If we instead use the minimum of
the observed color bimodality in the model galaxies (which is roughly
0.1 mags blueward of the location for the DEEP2 galaxies) our
qualitative results and conclusions remain unchanged.  
Details of each sample used in the mock
catalogs are given in Table 2.  The number densities for red
and blue galaxies in the model match the observed number densities quite well.

The resulting \xisp \ contours  for the semi-analytic model
galaxies are shown in Figure~\ref{mr_xisp}.  Red model galaxies are more
clustered and have stronger fingers of god than the blue model
galaxies.  A direct comparison between the red model galaxy \xisp \
and the equivalent DEEP2 result (Figure~\ref{xisp_all}) indicates that a)
the redshift-space distortions are generally similar between the model 
and data, including both the fingers of god and the coherent infall on large
scales, and that b) the overall clustering amplitude is larger in the
model; e.g. compare the thick $\xi=1$ lines.  
Hence, the discrepancy seen at $z=0$ appears to extend back to earlier epochs.  The
blue model galaxy \xisp \ shows differences with the data as well.  The left
panel of Figure~\ref{mr_xisp} shows coherent infall on large scales but no
small-scale redshift-space distortions, unlike the DEEP2 data.
This is reflected in the low value of $\sigma_{12}$ found for the model 
blue galaxies (see Table 2).

The discrepancies seen in Figure~\ref{mr_xisp} are highly significant, as seen in 
the projected correlation
functions shown in Figure~\ref{mr_compare}, where the observed DEEP2
results are shown for comparison.  On scales $r_p<0.5$ \mpch \ 
the blue cloud semi-analytic model correlation function is flat and
does not continue to rise as in the data.  In addition, in the
quasi-linear regime, $r_p=1-10$ \mpch, the blue model galaxies are 
less clustered than the data by $\sim 15$\%.  In contrast,
the red galaxies in the model are significantly more clustered than
the data, $\sim40$\% over most scales, as also seen at low redshift.
Power-law fits are given in Table 2, where slightly larger scales 
are used for the model fits ($r_p=0.4-20$ \mpch) than for the data, 
as the model blue galaxy \wprp \ is not a power law on smaller scales.

Finally, the red-blue cross-correlation function for the model galaxies is shown
in Figure~\ref{mr_cross} (solid line) along with the cross-correlation
function in the DEEP2 data (thin dashed line) and the geometric mean
of the red and blue auto-correlation functions in the model 
(dot-dash line).  The cross-correlation function in the model 
is clearly lower on small scales ($r_p<1$ \mpch) than either the
geometric mean or the DEEP2 data.  This reflects the lack of blue
galaxies around red galaxies in the model.
Interestingly, a hint of this same flattening of \wprp \ for blue 
galaxies in this model can be seen at $z=0$ in Figure 5 of 
\cite{Springel05b}; however they do
not plot below scales $r_p=0.5$ \mpch \ where for us the effect is most
pronounced.  

It appears that the discrepancies shown in Figures~\ref{mr_compare} and 
\ref{mr_cross} are due to the incorrect modeling of the
physics that determines the colors of satellite galaxies.  
If too few model satellite galaxies are blue compared to the
data at $z\sim1$, then both the model blue auto-correlation function and red-blue
cross-correlation function will have lower amplitudes on small scales
than seen in the DEEP2 data, inside the `one-halo' regime ($r_p<1$ \mpch).
 Similarly, an overabundance of red satellites on such
scales should also be present, as is seen in Figure~\ref{mr_compare}.

A similar effect is detected at low redshift when comparing the
fraction of blue SDSS galaxies in groups with these 
Millenium Run semi-analytic models. \cite{Weinmann06} find that the
blue fraction of galaxies in groups is much higher in the SDSS data 
than in the model, which predicts at $z=0.1$ that almost all satellite 
galaxies ($\sim$85\%) are red.

Our results provide a valuable way to test some of the physical
assumptions made by the semi-analytic model to better understand the
processes that may actually be occurring and their relative
importance.  We identify two overly simplified aspects of the model
satellite evolution that contribute to the clustering discrepancies 
shown in Figure~\ref{mr_compare}.  
The first is the way in which hot gas is stripped from halos
($M_{\rm halo} \gtrsim 10^{12} h^{-1} M_{\sun}$)
when the halo falls into the potential
of a more massive system.  As discussed by \cite{Weinmann06}, the
common assumption in most semi-analytic models is one of \emph{extreme
strangulation}, where all of the hot gas is instantly stripped and
added to that of the larger halo upon accretion.  This simplifying
assumption is important for galaxy colors because the dynamical
time for a sub-halo (and hence satellite galaxy) to merge is usually 
longer than the time it takes the stellar population to fade and
redden, which all model satellites will do in the absence of fresh fuel
for star formation.  This provides an overly-efficient channel for blue
satellites to transform into red satellites.

\cite{Weinmann06} point out that the \cite{Croton06} model does not
include a host halo mass dependence for the strangulation efficiency.  
Given that the ability of a parent halo to strip gas from a subhalo
likely depends on the density of the hot gas in the parent halo, 
a model which includes such a mass-dependent strangulation prescription
might provide a better match to the results shown here.

The second aspect of the model satellite evolution that may be discrepant
with the real universe is the assumption that local potential wells
(i.e. subhalos inside larger parent halos) 
cannot accrete gas, meaning that after a satellite
galaxy is captured by the larger halo it receives no new baryonic fuel
for later star formation from the intragroup or intracluster medium.
This may be generally true in N-body simulations which show that
subhalos tend to lose mass through tidal stripping as they spiral
inward, not gain mass.  However it is plausible that in the outer regions 
of a dark matter halo, the local potential of a
subhalo may dominate that of its host and attract new baryons from the
hot parent halo, baryons that may subsequently cool onto the
satellite.  Such gas would extend the star formation history of the 
satellite and hence modify the evoution of its color.

It is also important to note that there are processes that are
known to operate in cluster (and also group) environments that act to
remove cold gas from a satellite galaxy, such as tidal stripping and
harassment. These processes
are not included in the \cite{Croton06} semi-analytic model used here
(most semi-analytic models ignore such additional effects to
keep the models relatively simple).  Their inclusion would typically redden
the color of the satellite further with time, worsening an effect that is
already too efficient in the current model.  
Our work indicates that current prescriptions for satellite infall and its 
consequent effect on star formation may be too simplistic.  However, it 
also highlights the utility of
color-dependent clustering measurements in constraining such models.

\section{Summary and Discussion}

In this paper we use volume-limited subsamples of the DEEP2
Galaxy Redshift Survey data to measure the color and
luminosity dependence of galaxy clustering at $z\sim1$.  
We split the sample into blue, star-forming galaxies and red, nonstar-forming 
galaxies using the observed color bimodality in $(U-B)$.  We further
subdivide each sample into luminosity and finer color bins. 

In this section we first summarize our main results and then 
discuss their implications.  Our main findings are:

\begin{itemize}

\item  Red galaxies are much more strongly clustered than blue galaxies
at $z\sim1$, with a relative bias that is roughly as high as is found
locally ($b_{red}/b_{blue}\sim1.4-1.6$).  This implies that, for relatively bright galaxies, 
the color-density relation was as 
strongly in place 8 Gyr ago as it is observed to be today.

\item  The comoving clustering amplitudes that we measure for both blue and red 
galaxies are only slightly (10-20\%) lower than what is found locally. 
The measured values are somewhat higher than those in previous DEEP2 and VVDS 
results \citep{Coil04xisp,Meneux06}, though generally within 2$\sigma$.

\item  We find no statistically-significant dependence of galaxy clustering on luminosity
for either red or blue galaxies, within the luminosity range that we probe.
Additionally, within the red sequence there is no dependence of 
clustering on color, while within the blue cloud there is
a strong dependence on color.  The
color dependence and lack of luminosity dependence for both red and blue 
that we find is consistent with previous results at
$z=0.1$, for the luminosity range probed here.

\item The stronger dependence of clustering on luminosity found for all
galaxies in the DEEP2 sample \citep{Coil06lum} is likely due in part 
to the changing fraction of red versus blue galaxies as
a function of luminosity, as
we do not find as strong of a luminosity dependence here within either the
red or blue galaxy populations independently. 

\item The brightest blue galaxies have a significant rise in their 
correlation function on small scales ($r_p<0.2$ \mpch), which likely reflects
a lower satellite fraction compared to fainter blue galaxies; i.e., bright
blue galaxies are more likely to be central galaxies within their parent 
dark matter halos.  

\item The correlation between color and clustering within the blue cloud 
is likely a reflection of a correlation between stellar mass and halo mass.  
The redder galaxies in the blue cloud  have higher stellar 
masses \citep[see Figure 8 of ][]{CooperSFR} and are significantly more 
clustered than the bluer galaxies, which implies that they are in more massive
dark matter halos.

\item  Red galaxies show strong `fingers of god' at $z\sim1$,
indicating that they lie in virialized overdensities such as groups and clusters.
Blue galaxies show smaller `fingers of god'.

\item  Both blue and red galaxies show a flattening of \xisp \ on large
scales due to coherent infall of galaxies onto structures that are
still collapsing (the Kaiser effect).

\item  The projected correlation functions of brighter samples near
$L^*$ show deviations from a power law on small scales, within the
one-halo regime.  The deviations are more pronounced for blue galaxies
than for red galaxies.

\item  The red-blue cross-correlation function is consistent with the
geometric mean of the auto-correlation functions of blue and red
galaxies separately.  There is a marginal trend on small scales for
the 
cross-correlation to lie below the geometric mean.  Such a feature, if 
real, would indicate a lack of blue galaxies near red galaxies.
This could reflect a suppression of star formation in the centers of 
overdensities, such as groups, which are dominated by red galaxies at $z\sim1$ 
as seen in the group-galaxy cross-correlation function \citep{Coil05}.

\item  Green galaxies in the valley of the observed color bimodality 
display a similar large-scale clustering strength as red galaxies, but 
show a small-scale clustering amplitude and infall kinematics akin to blue 
galaxies.  The green galaxy population does not have strong `fingers of 
god' but does show coherent infall on large scales, and the correlation 
function has a similar slope to the blue galaxy population.  Green galaxies 
thus appear to be at the edges, but not in the centers, of the same 
overdensities as red galaxies.  
This is consistent with green galaxies being transition objects moving 
from the blue cloud to the red sequence and supports a picture of quenching 
happening preferentially on the outskirts of overdensities at $z\sim1$ 
on scales $r_p\sim0.5-2$ \mpch.

\item  We provide a direct test of environment and clustering statistics, 
comparing the relative bias as a function of color using a local galaxy
overdensity estimator with the two-point correlation function results, and
find good agreement between the two.

\item  Comparing with the semi-analytic galaxy evolution model of 
\cite{Croton06}, we find that red model galaxies are significantly more 
clustered than the DEEP2 galaxies, and that in the model the clustering
strength of blue galaxies is too low, especially on small scales ($r_p<0.5$ \mpch).  
These discrepancies are likely due to galaxies being 
quenched too efficiently in the model; i.e., star formation is shut down
in satellite galaxies too quickly.

\item These results all rely on precise, robust, spectroscopic redshifts.
The DEEP2 rms redshift
errors (determined from repeated observations) are $<35$ km s$^{-1}$, which is
unprecedented at $z\sim1$ and makes the \xisp \ analyses presented here 
possible.

\end{itemize}

From the observed bias of the main red and blue galaxy samples we can
infer the minimum dark matter mass for halos that host these galaxies.
Under the assumption that all galaxies are central galaxies and not
satellites (i.e., one galaxy per dark matter halo), the large scale
bias of the galaxy population can be matched to the bias of dark
matter halos using the formulae from e.g., Mo \& White (1996) or Sheth, Mo \&
Tormen (2001). \nocite{Mo96, Sheth01}
However, we know that not all of the DEEP2 galaxies are central
galaxies \citep{Gerke05,Conroy06,Zheng07}, and to obtain a more precise estimate of the
minimum dark matter halo mass we use the results presented in the
Appendix of \cite{Zheng07}, which take into account the inferred
satellite fraction from HOD modeling of the DEEP2
luminosity-dependent clustering results \citep{Coil06lum}.  At $z=0.9$,
the observed bias and number densities of the main red and blue galaxy
samples (with $M_B\le-20.0$) 
imply minimum dark matter halo masses of $M_{min}\sim2 \times 
10^{12} h^{-1} M_{\sun}$ for red galaxies and $M_{min}\sim5 \times 10^{11}
h^{-1} M_{\sun}$ for blue galaxies.  However, the halo occupation of
red and blue galaxies is likely complicated enough to warrant a full
HOD model, which we defer to a future paper \citep[see, e.g.,][for
modeling of local red and blue
samples]{vandenbosch03,Collister05,Zehavi05}.

\begin{figure}
\begin{center}
\scalebox{0.4}{\includegraphics{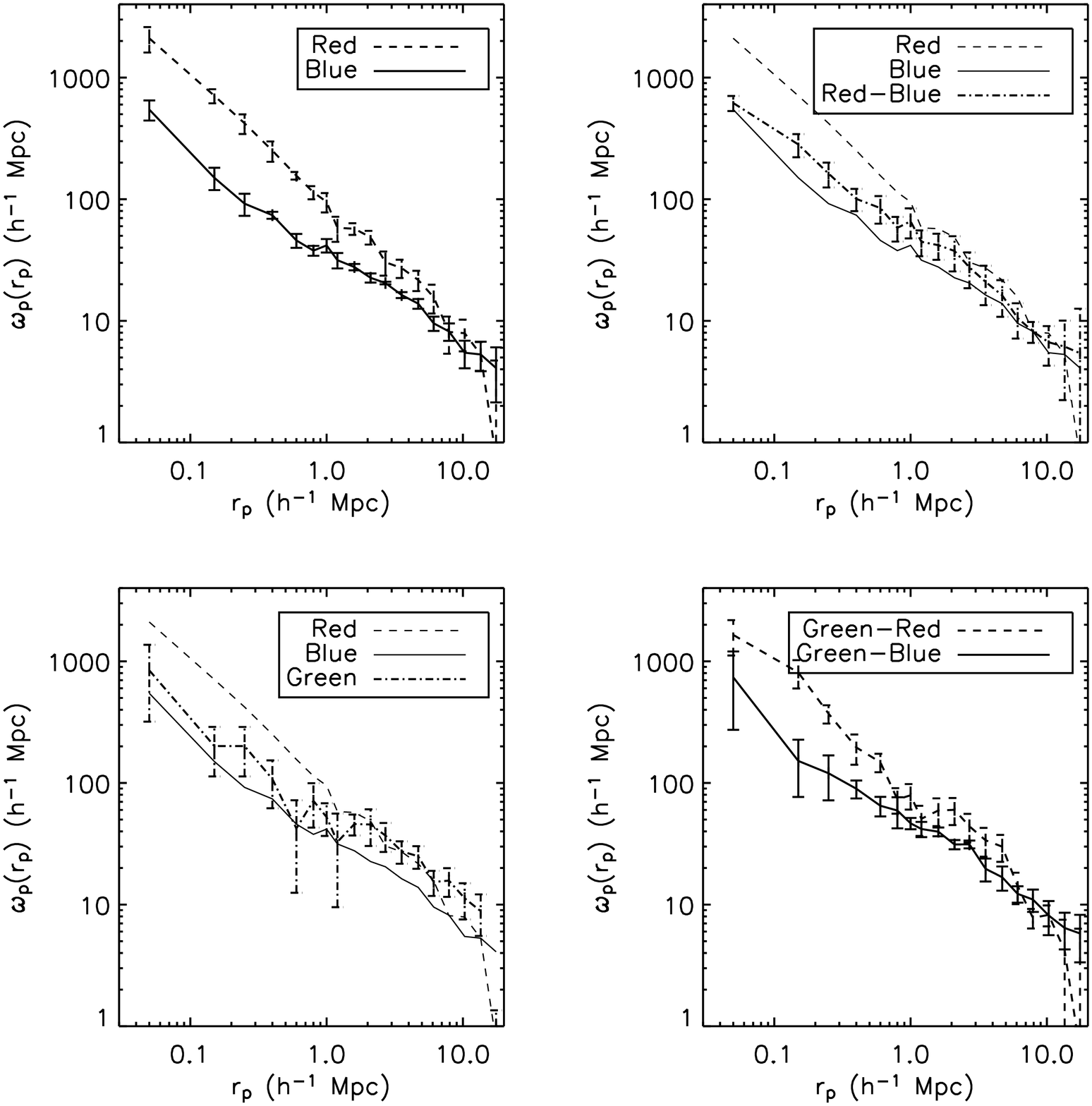}}
\end{center}
\caption{The projected correlation function, \wprp, for various color samples and cross-correlations.
Errors are computed from the variance across the ten pointings on the sky.  Corrections have been 
applied for observational biases due to slitmask effects (see text for details).
\label{wprp_all}}
\end{figure}

\begin{figure}
\begin{center}
\scalebox{0.4}{\includegraphics{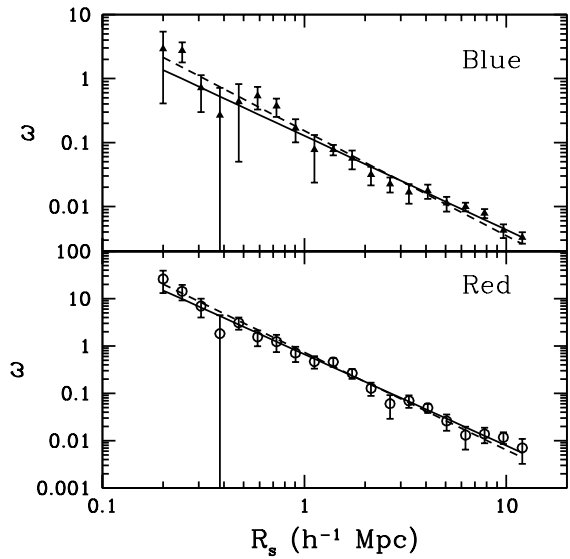}}
\end{center}
\caption{
The clustering of the main red and blue galaxy samples, as measured by the $\omega$
estimator of \cite{nikhil07} (see \S 3.3 for details).  Note that adjacent points are correlated. 
The solid line shows the 
best fit power law to these measurements, while the dashed line 
shows the result of a power-law fit to the \wprp \ points in Fig. 4.  
There is good agreement between the two fits, given the measurement errors.
\label{omega}}
\end{figure}

\begin{figure}
\begin{center}
\scalebox{0.5}{\includegraphics{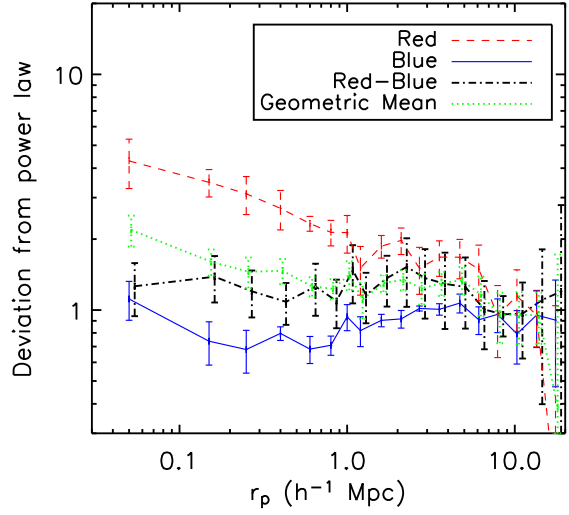}}
\end{center}
\caption{Deviations of \wprp \ from a power law with $r_0=4$ \mpch \ and $\gamma=1.8$ for the main
red and blue samples and the red-blue cross-correlation.  The cross-correlation is seen to be 
closer to the blue auto-correlation on small scales, at $r_p<1$ \mpch.  The geometric mean of the 
main red and blue \wprp \ is shown as a dotted line for comparison with the measured cross-correlation
function.
\label{cross_powerlaw}}
\end{figure}

\begin{figure}
\begin{center}
\scalebox{0.4}{\includegraphics{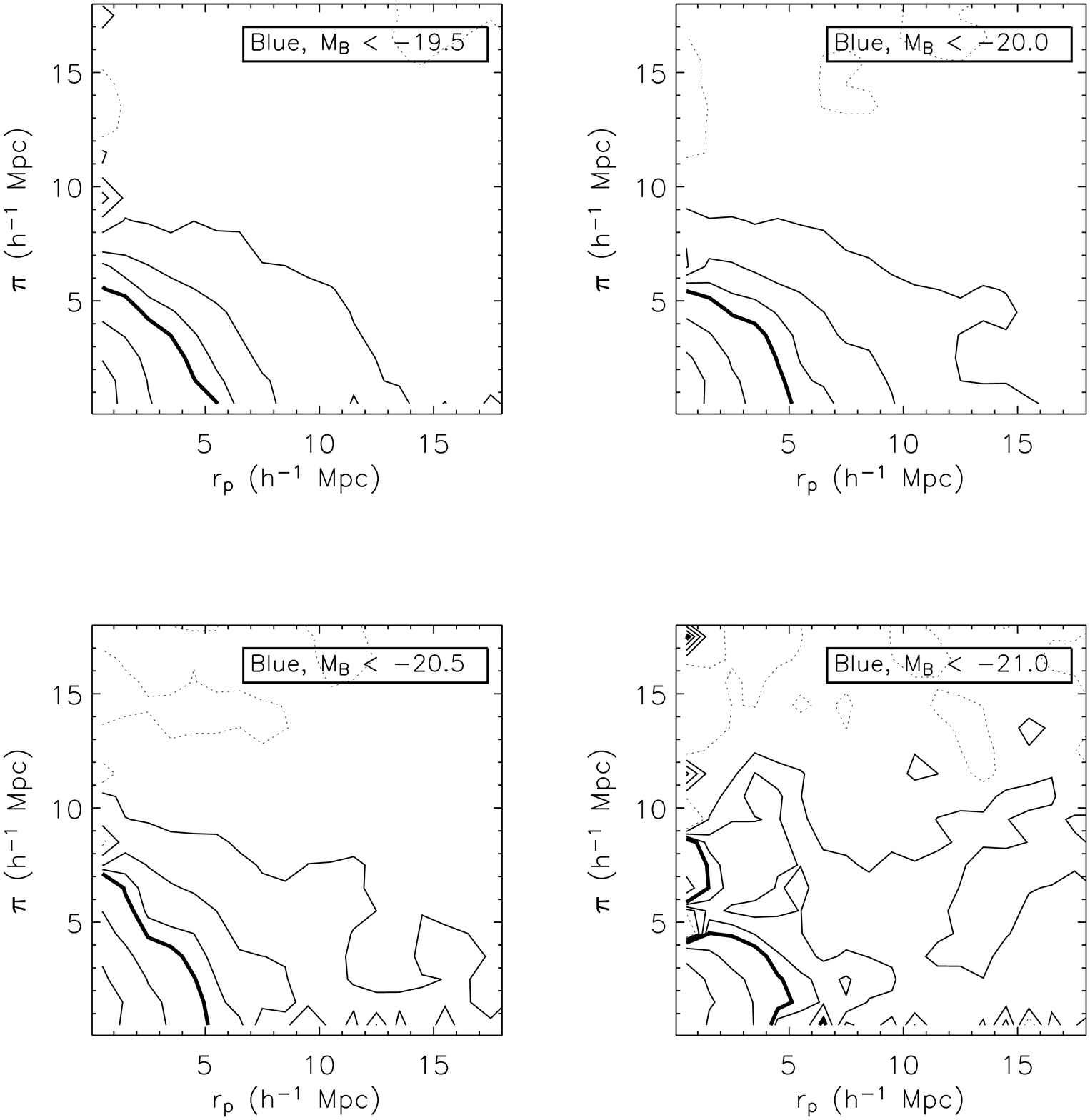}}
\end{center}
\caption{Contours of constant correlation strength 
for the two-dimensional correlation function, \xisp,
for the blue luminosity subsamples. A $1\times1$ \mpch \ boxcar smoothing has 
been applied in the figure, though it is not used in calculations.  Contour 
levels are the same as in Figure \ref{xisp_all}.
The strength of the fingers of god on small scales are seen to increase with 
luminosity in these samples.
\label{xisp_blue_lum}}
\end{figure}

\begin{figure}
\begin{center}
\scalebox{0.4}{\includegraphics{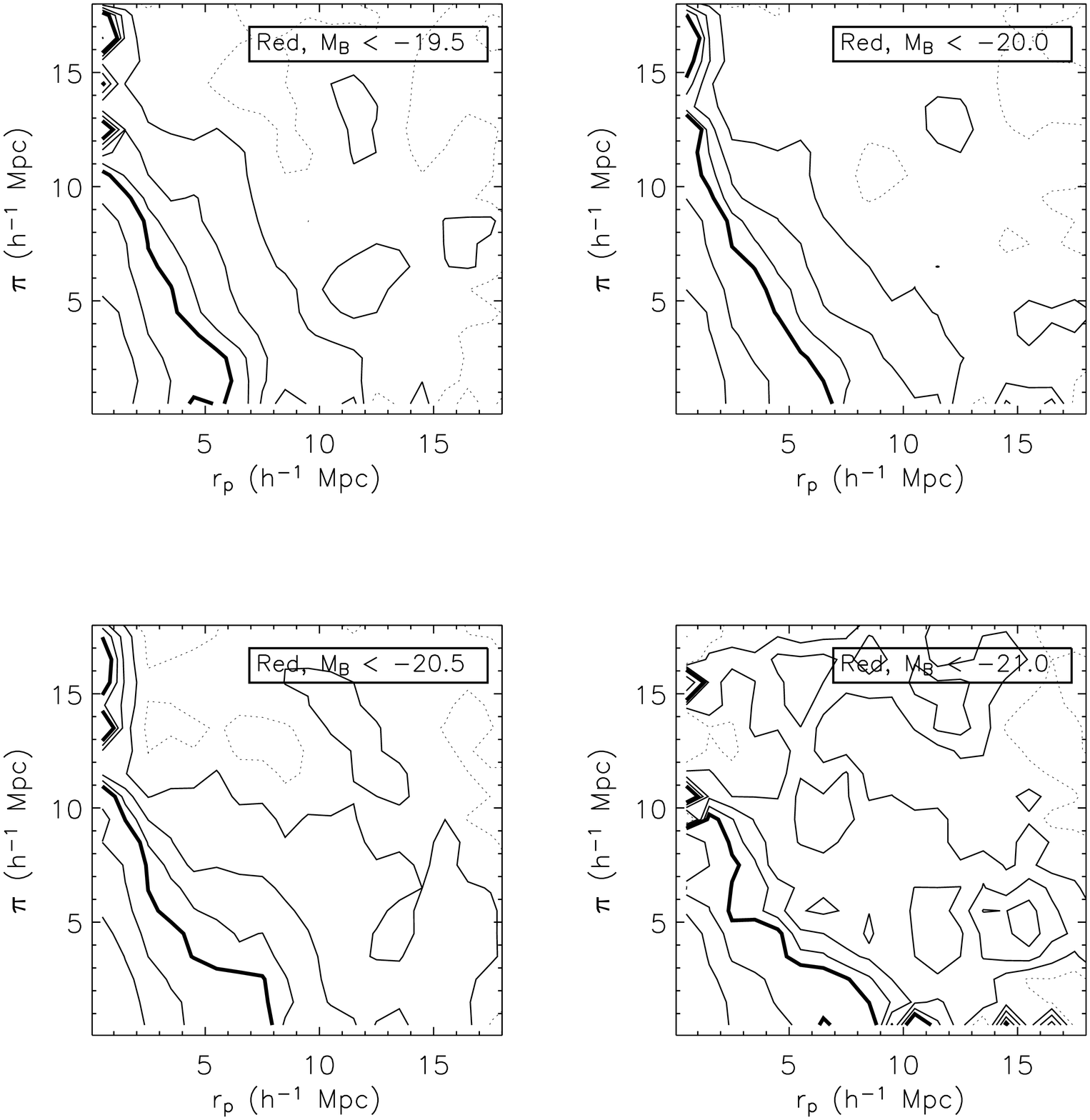}}
\end{center}
\caption{Contours of constant correlation strength 
for the two-dimensional correlation function, \xisp,
for the red luminosity subsamples. A $1\times1$ \mpch \ boxcar smoothing has been applied in the 
figure, though it is not used in calculations.  Contour levels are the same as in Figure \ref{xisp_all}.
Strong fingers of god are seen in each of the red luminosity samples.
\label{xisp_red_lum}}
\end{figure}

\begin{figure}
\begin{center}
\scalebox{0.4}{\includegraphics{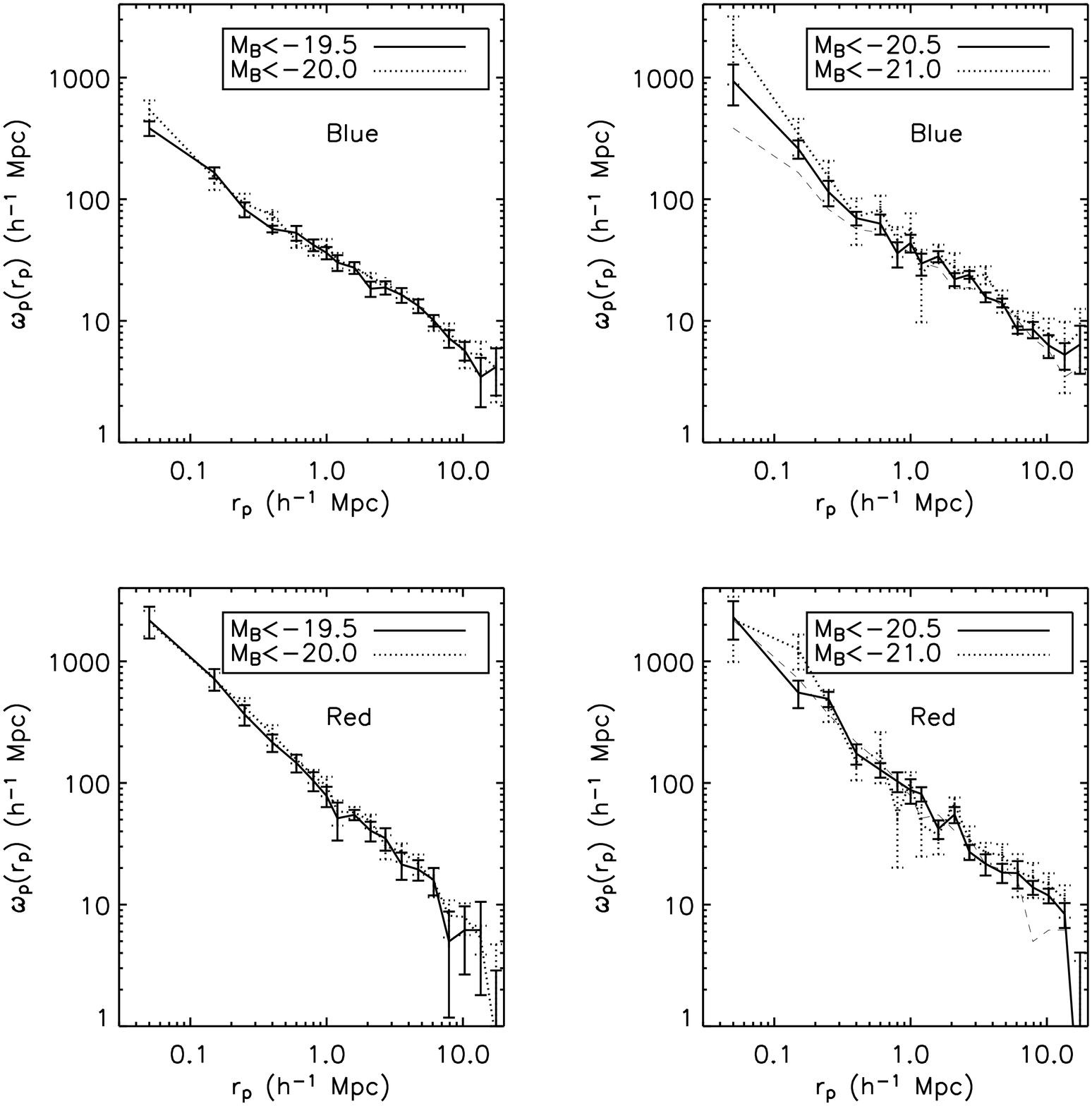}}
\end{center}
\caption{The projected correlation function, \wprp, for the blue (top ) and red (bottom) luminosity
subsamples. There is little luminosity dependence seen in the fainter samples (left).  Both blue and red 
galaxies show departures from a power law with a rise in the 
correlation function on small scales ($r_p<0.3$ \mpch) in the brighter samples with $M_B<-20.5$ (right).
In the right panels the thin dashed line shows the $M_B<-19.5$ results from the left panel for comparison.
\label{wprp_lum}}
\end{figure}

The different slopes measured for the clustering of red and blue
galaxies would pose a potential difficulty in terms of understanding how red
and blue galaxies populate dark matter halos, if the slopes were different
on large scales ($r_p\gtrsim2$ \mpch).  On these scales the halo bias is 
linear, in that there is an amplitude offset between \xir \ for halos of 
different mass, but no difference in the slope \citep[e.g.,][]{vandenbosch03, Conroy06}.  
We do not have a large enough survey here to claim a significant difference in
the slope on scales $r_p\gtrsim2$ \mpch, though the upper left panel
of Figure \ref{wprp_all} does suggest such a difference.

It has been previously established that the number density of galaxies
on the red sequence has increased since $z=1$
\citep{Bell04,Faber06,Brown07}.  It is thought that these galaxies 
likely migrated from the blue cloud to the red sequence as their star 
formation was quenched.  Using the observed clustering of red and blue 
galaxies at $z\sim1$, one can predict what 
their clustering strength should be at $z\sim0$, for a given cosmology and
assuming no merging of galaxies, by using the continuity equation and the 
expected growth of perturbations \citep{Tegmark98}.
This leads to larger predicted clustering amplitudes than is 
observed for local $L^*$ red and blue galaxies; the main red and 
blue samples used here (both of which have $L\sim L^*$ at $z=0.9$) 
would have clustering scale-lengths of $r_0=6.5$ \mpch \ and $r_0=5.5$ \mpch, 
respectively, at $z=0$.  The fact that local clustering measurements at $L^*$ are
lower than these predictions, especially for blue galaxies, lends credence to the 
idea that blue galaxies have migrated to the red sequence and less clustered blue 
galaxies have joined the blue cloud since $z=1$.

The color and luminosity dependence of clustering at $z=1$
can illuminate {\it which} blue galaxies likely become red between
$z=1$ and $z=0$.  The fact that the relative bias between red and blue 
galaxies is as high at $z\sim1$ as $z\sim0$ indicates that blue galaxies 
have to be turning red {\it in a density-dependent way}, 
in that the more clustered of the blue galaxies must become red.
The bias for any galaxy population should tend towards unity with time 
\citep{Fry96} and linear growth theory would predict that the relative bias 
of our red and blue samples should decrease by 8\% to $z=0.1$.  
Observationally, however, it appears that the relative bias between
red and blue galaxies has either remained constant since $z\sim1$ or 
increased.  This therefore implies that the most clustered blue 
galaxies must be turning red to boost the clustering of red galaxies relative to 
blue galaxies and to counteract the effect of the relative bias decreasing.
It is the most clustered of the blue galaxies at
$z=1$ that are likely to be as clustered as red galaxies by $z=0$, and our 
results here indicate that the reddest of the blue galaxies and
the green galaxies are the most likely candidates for having their
star formation quenched by $z=0$, based on their clustering
properties.

The luminosity dependence of clustering at $z\sim1$ can be explained as
being almost entirely due to the relationship between luminosity and halo mass,
 as shown by \cite{Conroy06} who
use a simple one-to-one relation between halo mass and light to obtain the
same luminosity- and scale-dependence of clustering in an N-body
simulation as we find in DEEP2.  The tight relation
between halo mass and galaxy light is confirmed by \cite{Zheng07}, who
fit HOD models to the DEEP2 luminosity dependent clustering results.
However, the color dependence of clustering seen here can not entirely be
halo-mass dependent, as the clustering as a function of color does not match
the results of \cite{Conroy06}, where the galaxies have been ranked by 
luminosity, which is equivalent to halo mass; i.e., the trends seen with 
color in the data do not match the trends found with luminosity or halo mass 
in the model.
This implies that there is a wider range in the $M_{halo}/L$ ratio of galaxies 
at a given color than at a given luminosity.  To constrain galaxy evolution 
physics it may therefore be more informative to study samples divided by 
color rather than 
luminosity, as more galaxy physics is apparently needed to interpret the
color dependence of galaxy clustering.  The lack of a 
stronger luminosity dependence for the separate red and blue galaxy samples 
shown here makes the results of \cite{Conroy06} more surprising, as it is 
not clear why $M_B$ luminosity for all galaxies should be so tightly 
correlated with halo mass.  It may be that the relative mix
of red and blue galaxies as a function of luminosity helps to create a 
stronger relation with luminosity for all galaxies and is more tightly 
correlated with halo mass than the luminosity itself.

As noted in previous papers on the color-density relation in the
DEEP2 sample \citep{Gerke07,Cooper06}, the observed color-density
relation at $z\sim1$ can not be the result of cluster-specific physics
as only a few percent of DEEP2 galaxies are in massive clusters.  The
color dependence of clustering must be dominated by either processes
that operate in a group environment or intrinsic processes that depend
on the age, stellar mass or halo mass of a galaxy.  
Processes that occur when a galaxy is accreted into a larger halo, becoming
a satellite galaxy, may {\it contribute} to the observed color-density relation
but do not {\it dominate} it, as most DEEP2 galaxies are not satellite galaxies 
\citep{Gerke05,Zheng07}. What 
star-formation quenching processes may be at work in groups?  There
are more galaxy mergers in groups than in clusters, such that mergers
may play an important role in shutting off star formation in group
galaxies.  There is less hot ambient gas in groups than in clusters;
it is not clear whether the density of the intragroup gas is high
enough to strip gas from infalling satellite galaxies, except near the
core of the group.  Apparently not all satellite galaxies are red, as
is seen in the clustering results; this implies that there is a
non-negligible timescale for quenching star formation in satellite
galaxies.
 
\begin{figure}
\begin{center}
\scalebox{0.5}{\includegraphics{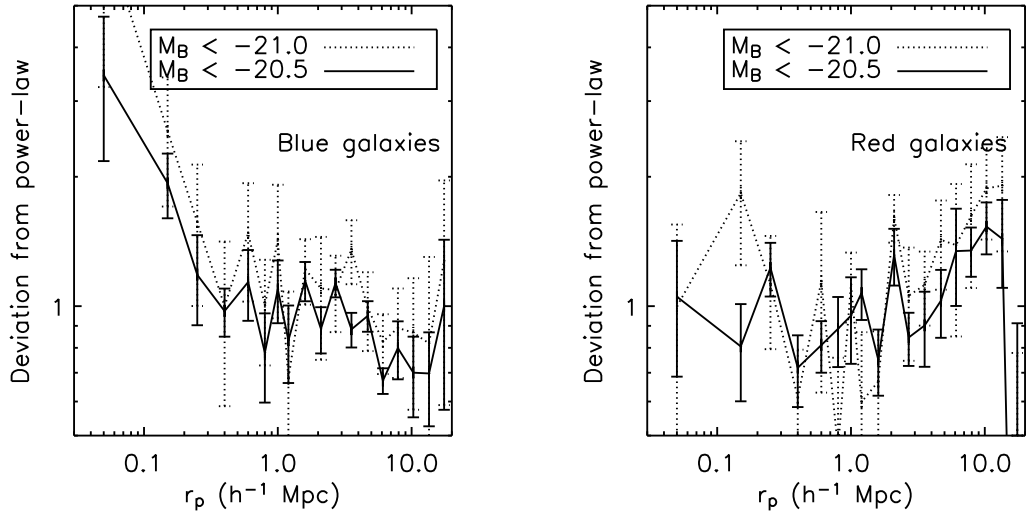}}
\end{center}
\caption{Deviations of \wprp \ from the best fit power law for the main blue and red samples for 
the brighter blue (left) and red (right) luminosity samples.  A significant rise is seen on small
scales for both blue and red galaxies in the brightest samples.
\label{wprp_lum_powerlaw}}
\end{figure}

\begin{figure}
\begin{center}
\scalebox{0.4}{\includegraphics{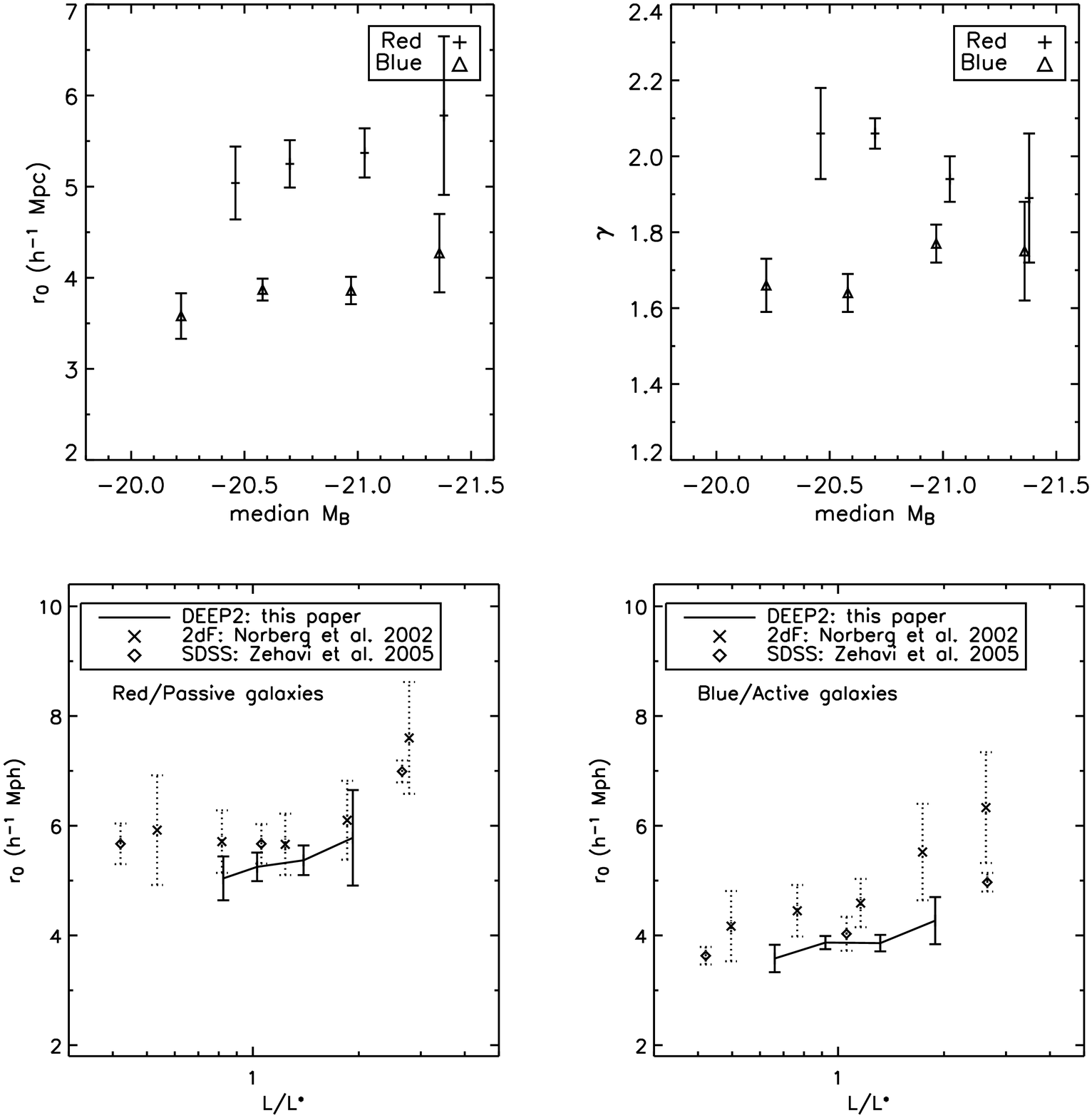}}
\end{center}
\caption{The clustering scale-length \rr \ (left) and slope 
$\gamma$ (right) for DEEP2 galaxies as a function of 
 median absolute $M_B$ magnitude (top) and $L/L^*$ (bottom) for both
red and blue galaxies. The values of \rr \ and 
$\gamma$ for each luminosity sample are given in Table 2.
 Errors are estimated using the variance of power-law fits to
 jacknife samples.  In the lower panels we compare the scale-length 
for red (left) and blue (right) galaxies found here at $z\sim1$ with
local results from 2dF \citep{Norberg02} and SDSS \citep{Zehavi05}.
\label{r0_gam_lum}}
\end{figure}

\begin{figure}
\begin{center}
\scalebox{0.5}{\includegraphics{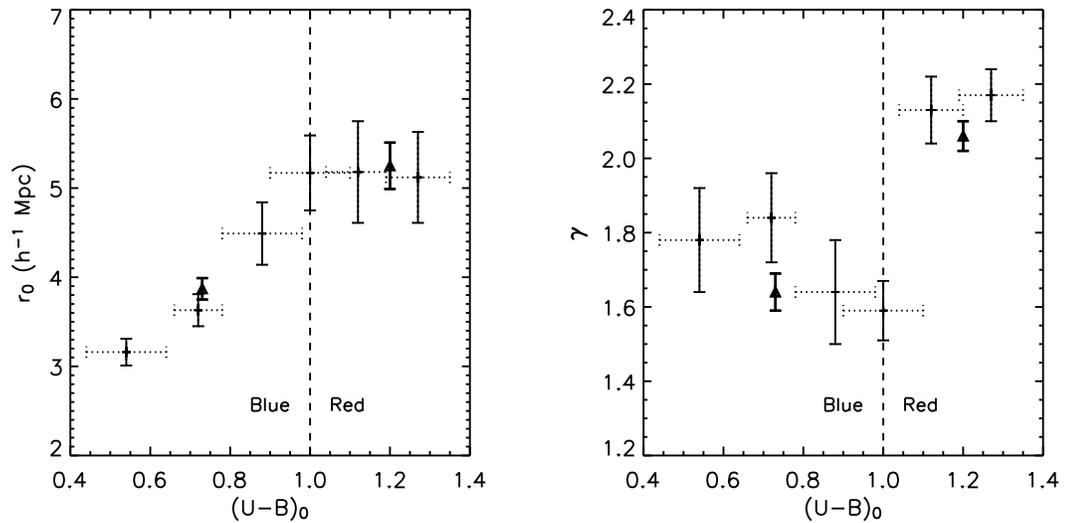}}
\end{center}
\caption{The clustering scale-length, \rr \ (left), and slope, 
$\gamma$ (right), of DEEP2 galaxies as a function of 
 restframe color. The values of \rr \ and 
$\gamma$ for each color sample are given in Table 2.
 Errors are estimated using the variance of power-law fits to
 jacknife samples.  Dotted horizontal lines
 indicate the color range for each point; for clarity the color
 ranges for the main blue and red samples, shown as triangles, are
 omitted.  The vertical dashed line indicates the approximate
 location of the minimum in the color bi-modality which separates red
 and blue galaxies.
\label{r0_gam_color}}
\end{figure}

\begin{figure}
\begin{center}
\scalebox{0.5}{\includegraphics{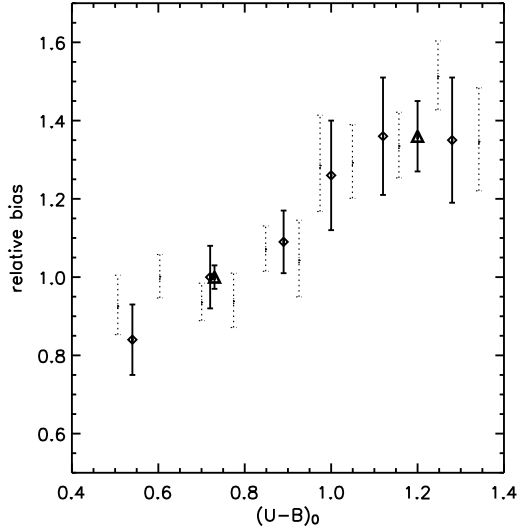}}
\end{center}
\caption{The mean relative bias of each of the color samples compared
to the main blue sample, averaged on scales $r_p=1-5$ \mpch. 
Triangles show the main red and blue samples and diamonds show the 
finer color bins.  
 Dotted lines show the relative overdensity as 
a function of color derived
from the $\delta_3$ environment parameter of \cite{Cooper06}, again
normalized to the main blue sample color, which agrees very well
with the relative bias measured from the correlation function.
\label{relbias}}
\end{figure}

\begin{figure}
\begin{center}
\scalebox{0.5}{\includegraphics{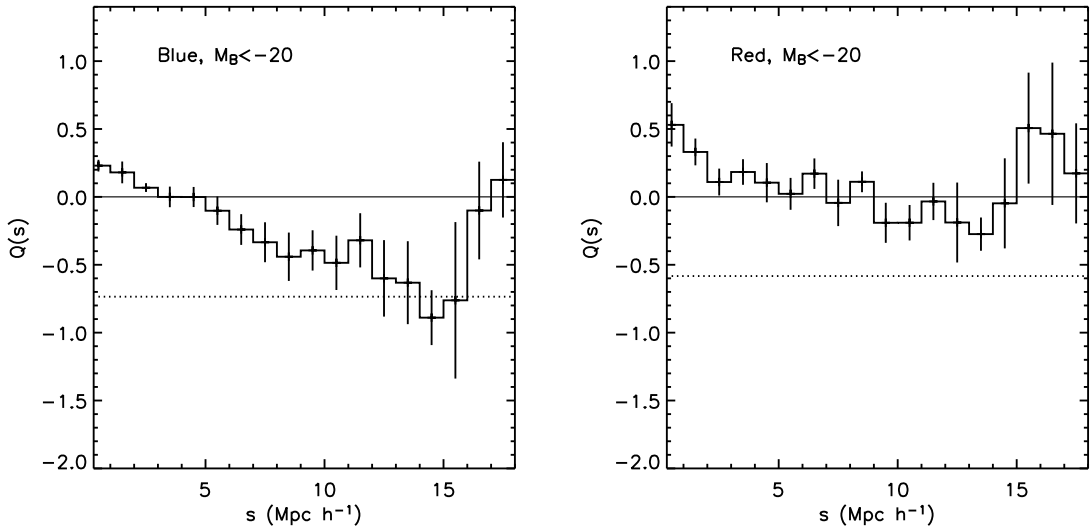}}
\end{center}
\caption{The quadrupole-to-monopole ratio, defined as
$Q(s)\equiv\xi_2/(\overline{\xi_0}-\xi_0)$, as a function of scale for
the main blue (left) and red (right) samples.  Dotted lines show the
expectations from linear theory for $\beta=0.65$ for blue galaxies and
$\beta=0.52$ for red galaxies, where $\beta\equiv\Omega_M^{0.6}/b$ and
$b$ is the linear galaxy bias.
\label{qratio}}
\end{figure}

\begin{figure}
\begin{center}
\scalebox{0.5}{\includegraphics{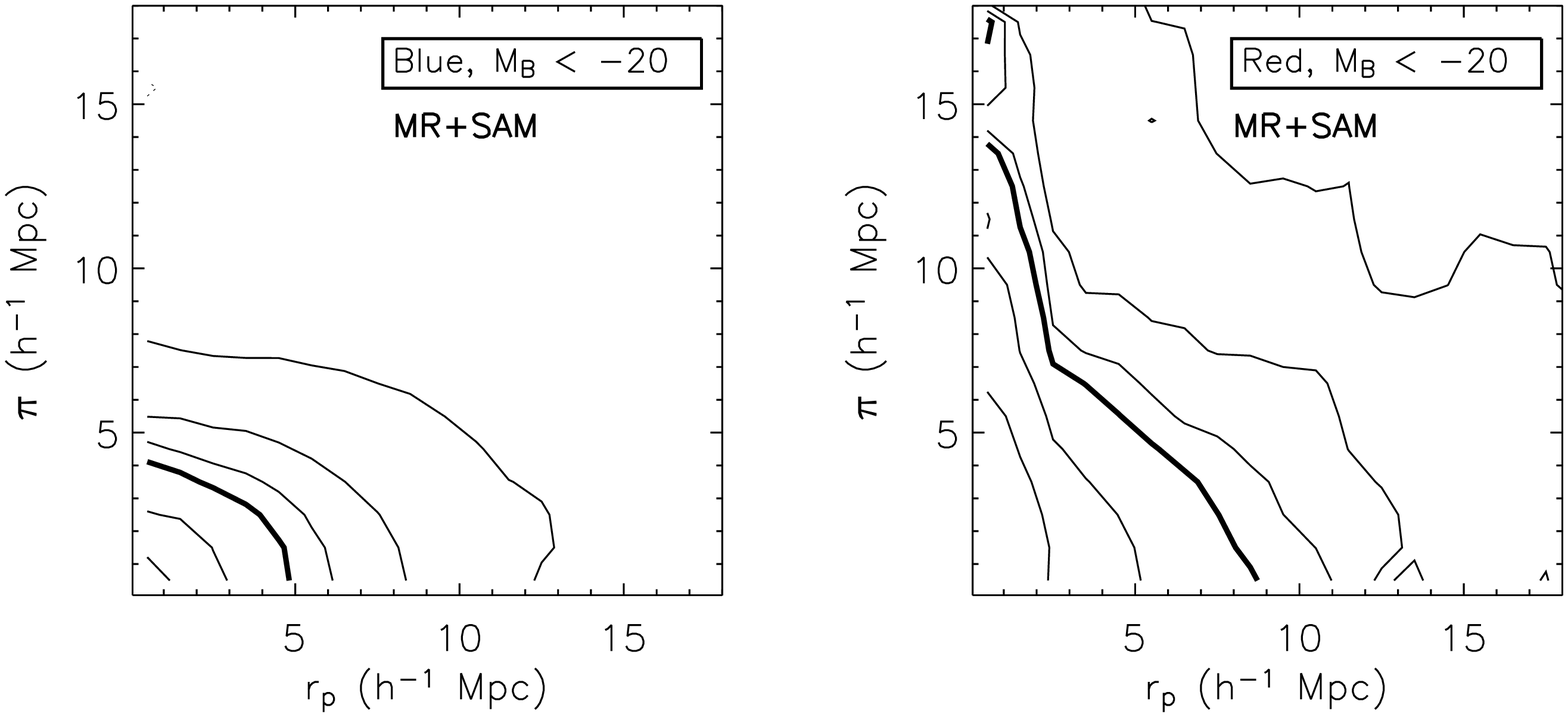}}
\end{center}
\caption{
The two-dimensional correlation function, \xisp, for blue (left) and
red (right) galaxies in the Millenium Run semi-analytic model 
mock galaxy catalogs \citep{Croton06} matched to the DEEP2 survey.  A $1\times1$
\mpch \ boxcar smoothing has been applied in the figure, and contours
levels are the same as in Figure \ref{xisp_all}.  Almost no small scale 
redshift-space distortions are seen for blue galaxies, while red galaxies show
strong fingers of god as well as coherent infall on large scales.
\label{mr_xisp}}
\end{figure}

\begin{figure}
\begin{center}
\scalebox{0.5}{\includegraphics{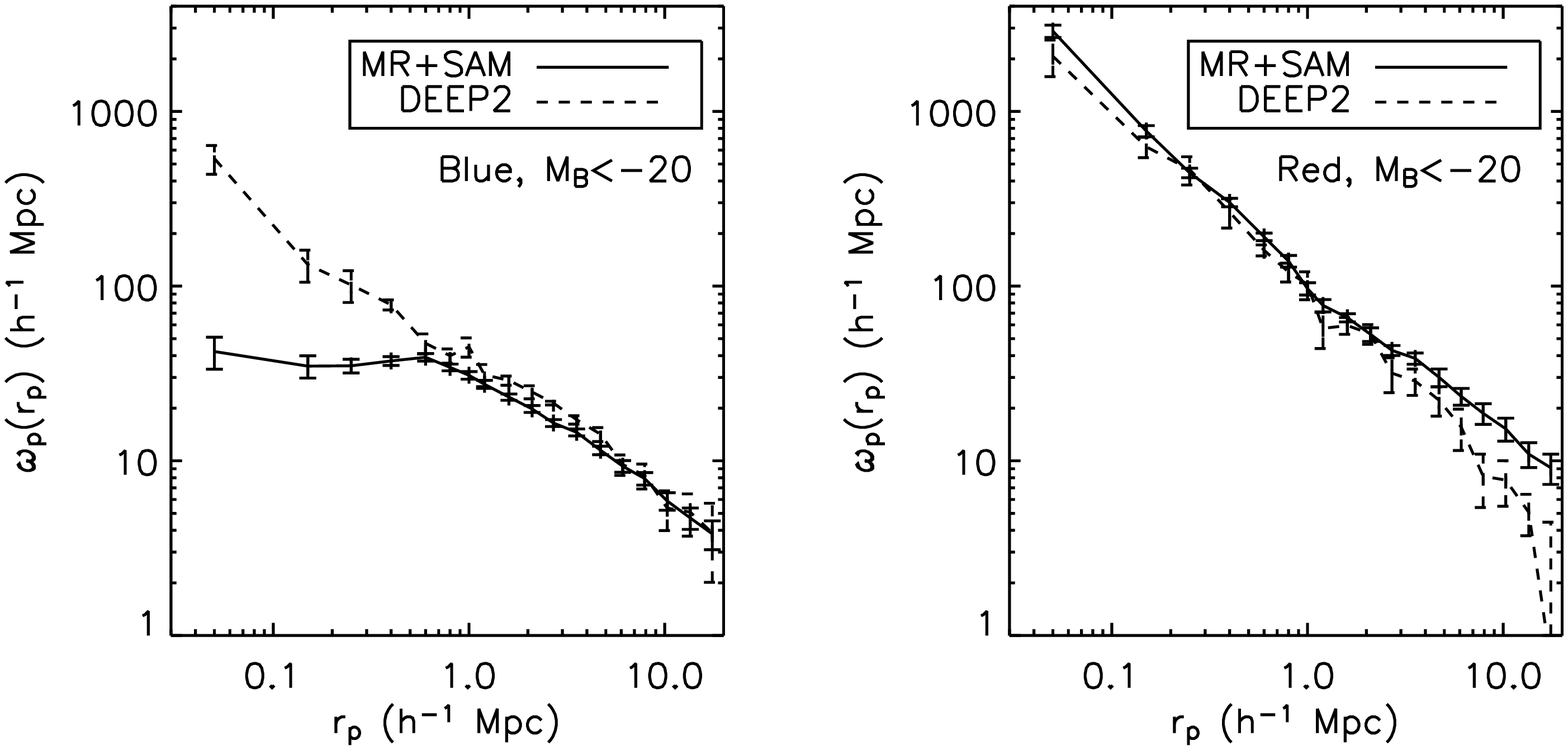}}
\end{center}
\caption{
The projected correlation function, \wprp, for blue (left) and red
(right) galaxies in the  Millenium Run  semi-analytic model  
mock catalogs  (solid line) and the DEEP2 galaxy data (dashed line).
There is a strong discrepancy between the model and the data
for blue galaxies on small scales ($r_p<1$ \mpch), where the model 
does not have enough blue galaxies.  The red galaxy correlation
function is too high on all scales in the model compared to
the data.
\label{mr_compare}}
\end{figure}

\begin{figure}
\begin{center}
\scalebox{0.5}{\includegraphics{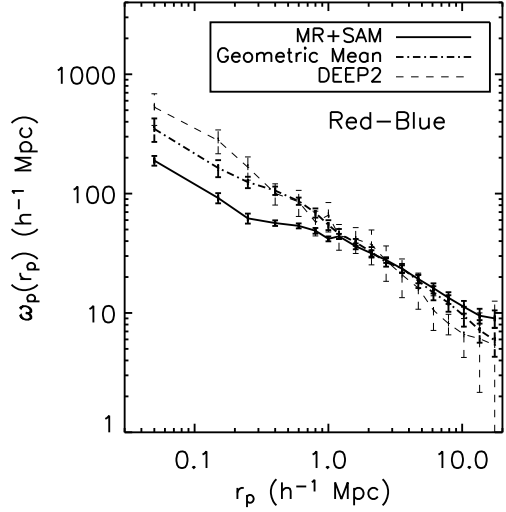}}
\end{center}
\caption{
The projected cross-correlation function for blue and red galaxies in
 the Millenium Run semi-analytic model mock catalogs
(solid line) and the DEEP2 galaxy data (thin dashed line).  
The geometric mean of the blue and red auto-correlation functions in the
semi-analytic model is shown for comparison (dot-dash line).
On small scales ($r_p<1$ \mpch) the cross-correlation function in the 
model is significantly lower than both the geometric mean of
the blue and red galaxies in the model and the
cross-correlation function in the DEEP2 data.
\label{mr_cross}}
\end{figure}

DEEP2 galaxies in groups at $z\sim1$ have a clustering scale-length of 
$r_0=4.97 \pm0.25$ \mpch \ and a slope of $\gamma=2.15 \pm0.06$ \citep{Coil05}, 
similar to red galaxies as measured here.  Galaxies in the field at $z\sim1$
have a slope of $\gamma=1.70 \pm0.13$, similar to blue galaxies, but
with a lower correlation length, $r_0=2.54 \pm0.25$ \mpch.  
The cross-correlation of red galaxies and groups is higher than 
that of blue galaxies and groups \citep{Coil05}, which also indicates 
that red galaxies are more likely to be in groups. However, the picture 
is not as simple as all red galaxies being
in groups and blue galaxies being in the field: blue galaxies have 
some fingers of god on small scales, indicating that at least some are in virialized 
overdensities, and the brighter blue galaxies show a rise in
the correlation function on small scales, which is probably due to a lower
satellite fraction for the brighter blue samples.  These galaxies are likely
the central galaxies in groups at $z=1$, and may migrate
to the red sequence by $z\sim0$.  The scale at which the one-halo term rise
occurs for these samples is small, $r_p\sim 0.3$ \mpch \ 
(as seen in Figure \ref{wprp_lum_powerlaw}), and indicates again that these
galaxies are likely to be in the centers of the groups they reside in.
Additionally, $\sim5-10$\% of DEEP2 galaxies in underdense regions are red
\citep{Cooper07}.

The green galaxy population, which lies near the minimum in the
observed color bimodality, appears to be, at least in part, 
a distinct population and is
not a simple mix of red and blue galaxies.  Though the green population
can include the tail of each of the blue and red distributions, it also
contains transitional objects, likely moving from the blue cloud to the
red sequence.  
The green galaxies have different clustering characteristics than either the
red or blue galaxies as a whole.  On large scales ($r_p>1$ \mpch) the green
galaxies are as clustered as red galaxies of the same luminosity, implying
that they reside in the same relatively massive halos as red galaxies.
On small scales, however, the clustering amplitude is similar to that 
of blue galaxies, which likely reflects a lower radial concentration 
for green galaxies compared to red galaxies.  
The general picture that emerges is
one in which star formation quenching is occurring primarily at the
edges of overdensities as galaxies fall in, turning blue galaxies
green around the edges of groups.  

We have identified that the scale at which this transition from red to blue
occurs at $z\sim1$ is 0.5 \mpch \ $<r_p<$ 2 \mpch.  This scale is just 
larger than the transition scale between the one-halo and two-halo terms 
at $z\sim1$, which is shown by \cite{Zheng07} (see their Fig. 1) to be 
$\sim0.5$ \mpch \ for all galaxies and by \cite{Coil05} (see their Fig. 8) to
be $\sim1.0$ \mpch \ for galaxies in groups.  This transition scale for 
green galaxies is therefore similar to or just larger than the largest halos, 
which indicates that it is occurring at $\sim1-2$ virial radii, on the 
edges of large halos.  This conclusion lends support to the idea that the 
mechanism which shuts off star formation in these galaxies is not happening
at the centers of massive halos, but on the outskirts.

The comparison to semi-analytic galaxy formation models demonstrates vividly the
utility of using correlation statistics to test theoretical models, and
the need to test them at intermediate redshift, not just locally.  It
is also likely that currently popular models that have a simple halo
threshold mass for quenching star formation in galaxies will suffer similar
problems as the model shown here. 
We urge theorists to use these data to constrain their models.

\acknowledgements

We thank Charlie Conroy, Daniel Eisenstein, Jeremy Tinker, 
Frank van den Bosch, Christopher
Willmer and Zheng Zheng for useful discussions and comments on this paper.
ALC and NP are supported by NASA through Hubble Fellowship grants
HF-01182.01-A and HST-HF-01200, respectively, awarded by the Space
Telescope Science Institute, which is operated by the Association of
Universities for Research in Astronomy, Inc., for NASA, under contract
NAS 5-26555.
The DEIMOS spectrograph was funded by a grant from CARA (Keck
Observatory), an NSF Facilities and Infrastructure grant (AST92-2540),
the Center for Particle Astrophysics and by gifts from Sun
Microsystems and the Quantum Corporation. 
The data presented herein were obtained at the W.M. Keck Observatory, which is
operated as a scientific partnership among the California Institute of
Technology, the University of California and the National Aeronautics
and Space Administration. The Observatory was made possible by the
generous financial support of the W.M. Keck Foundation. The DEEP2 team
and Keck Observatory acknowledge the very significant cultural role
and reverence that the summit of Mauna Kea has always had within the
indigenous Hawaiian community and appreciate the opportunity to
conduct observations from this mountain.


\clearpage
\pagestyle{empty}
\begin{deluxetable}{llllllll}
\tablewidth{0pt}
\tablecaption{Published Color Clustering Results}
\tablehead{
\colhead{}&\colhead{mean}&\colhead{blue}&\colhead{blue}&\colhead{red}&\colhead{red}&\colhead{relative}&\colhead{M}\\
\colhead{}&\colhead{$z$}&\colhead{$r_0$}&\colhead{$\gamma$}&\colhead{$r_0$}&\colhead{$\gamma$}&\colhead{bias}&\colhead{range}\\
}
\startdata
Norberg et al. 2002  & 0.1 & $4.45 \pm0.47$ & $1.76 \pm0.09$    & $5.71 \pm0.57$ & $1.87 \pm0.09$ & -- &        $-20<M_{b_J}<-19$ \\
Madgwick et al. 2003 & 0.1 & $3.67 \pm0.30$ & $1.60 \pm0.04$    & $6.10 \pm0.34$ & $1.95 \pm0.03$ & $1.45 \pm0.14$ & $\sim M^*$ \\
Budavari et al. 2003 & 0.2 & $\sim$4.5-6.0  & $\sim1.67 \pm0.10$& $\sim6.3-6.6 \pm0.20$ & $2.0 \pm0.5$ & -- &   $-20>M_{r*}>-21$ \\
Zehavi et al. 2004   & 0.1 & $3.63 \pm0.16$ & $1.69 \pm0.04$    & $5.67 \pm0.37$ & $2.08 \pm0.05$ &  $\sim1.6$ &       $-20<M_r<-19$ \\
Coil et al. 2004     & 0.8 & $2.81 \pm0.48$ & $1.52 \pm0.06$    & $4.32 \pm0.73$ & $1.84 \pm0.07$ & $1.41 \pm0.10$ & \\ 
Meneux et al. 2006   & 0.8 & $2.18 \pm0.30$ & $1.40 \pm0.14$    & $3.78 \pm0.7$  & $1.87 \pm0.25$ & $1.45 \pm0.27$ & \\
\enddata
\label{published}
\end{deluxetable}

\begin{deluxetable}{lrcccclcccccl}
\tabletypesize{\footnotesize}
\tablewidth{0pt}
\tablecaption{DEEP2 Galaxy Samples}
\tablehead{
\colhead{Sample}&\colhead{No. of}&\colhead{$n$}&\colhead{median}&\colhead{$M_B$}&\colhead{Median}&\colhead{$ \ \ \ z$}&\colhead{Mean}&\colhead{$r_0$}&\colhead{$\gamma$}&\colhead{Bias\tablenotemark{a}}&\colhead{$\ \ \ \sigma_{12}$}\\
\colhead{}      &\colhead{galaxies}&\colhead{$(h^3 Mpc^{-3}$)}&\colhead{$(U-B)$}&\colhead{range}&\colhead{$M_B$} &\colhead{range}&\colhead{$z$} &               &                  &\colhead{}&
}
\startdata
Red: -21.0  &   792         & $3.9 \ 10^{-4}$ & $1.22$          & $\le-21.0$  &  $-21.38$  & $0.7-1.1$   & $0.92$ & $5.78 \pm0.87$ & $1.89 \pm0.17$ & $1.79 \pm0.21$ & $360 \pm70$ \\ %
Red: -20.5  &  1370         & $9.0 \ 10^{-4}$ & $1.21$          & $\le-20.5$  &  $-21.03$  & $0.7-1.025$ & $0.88$ & $5.37 \pm0.27$ & $1.94 \pm0.06$ & $1.71 \pm0.14$ & $500 \pm50$ \\ %
Red: -20.0 (main)  &  1474  & $1.6 \ 10^{-3}$ & $1.21$          & $\le-20.0$  &  $-20.70$  & $0.7-0.925$ & $0.82$ & $5.25 \pm0.26$ & $2.06 \pm0.04$ & $1.65 \pm0.15$ & $530 \pm50$ \\ %
Red: -19.5  &  1032         & $2.4 \ 10^{-3}$ & $1.20$          & $\le-19.5$  &  $-20.46$  & $0.7-0.825$ & $0.77$ & $5.04 \pm0.40$ & $2.06 \pm0.12$ & $1.55 \pm0.15$ & $490 \pm40$ \\ %
Blue: -21.0 &  1424         & $6.1 \ 10^{-4}$ & $0.81$          & $\le-21.0$  &  $-21.28$  & $0.7-1.2$   & $1.00$ & $4.27 \pm0.43$ & $1.75 \pm0.13$ & $1.39 \pm0.16$ & $250 \pm50$ \\ %
Blue: -20.5 &  3747         & $1.6 \ 10^{-3}$ & $0.77$          & $\le-20.5$  &  $-20.89$  & $0.7-1.2$   & $0.99$ & $3.86 \pm0.15$ & $1.77 \pm0.05$ & $1.30 \pm0.07$ & $360 \pm30$ \\ %
Blue: -20.0 (main) &  4808  & $3.5 \ 10^{-3}$ & $0.74$          & $\le-20.0$  &  $-20.48$  & $0.7-1.05$  & $0.90$ & $3.87 \pm0.12$ & $1.64 \pm0.05$ & $1.28 \pm0.04$ & $240 \pm20$ \\ %
Blue: -19.5 &  4018         & $6.3 \ 10^{-3}$ & $0.68$          & $\le-19.5$  &  $-20.10$  & $0.7-0.9$   & $0.81$ & $3.58 \pm0.25$ & $1.66 \pm0.07$ & $1.24 \pm0.08$ & $270 \pm20$ \\ %
Red: Redder &   731         & $7.9 \ 10^{-4}$ & $1.27$          & $\le-20.0$  &  $-20.66$  & $0.7-0.925$ & $0.81$ & $5.12 \pm0.51$ & $2.17 \pm0.07$ & $1.60 \pm0.21$ & $740 \pm180$ \\ %
Red: Bluer  &   743         & $8.1 \ 10^{-4}$ & $1.12$          & $\le-20.0$  &  $-20.61$  & $0.7-0.925$ & $0.82$ & $5.18 \pm0.57$ & $2.13 \pm0.09$ & $1.61 \pm0.22$ & $740 \pm190$ \\ %
Green       &   997         & $7.7 \ 10^{-4}$ & $1.00$          & $\le-20.0$  &  $-20.65$  & $0.7-1.0$   & $0.87$ & $5.17 \pm0.42$ & $1.59 \pm0.08$ & $1.65 \pm0.26$ & $490 \pm110$ \\ %
Blue: Redder&  1407         & $1.3 \ 10^{-3}$ & $0.88$          & $\le-20.0$  &  $-20.65$  & $0.7-1.0$   & $0.88$ & $4.49 \pm0.35$ & $1.64 \pm0.14$ & $1.44 \pm0.19$ & $320 \pm40$ \\ %
Blue: Middle&  1204         & $1.1 \ 10^{-3}$ & $0.72$          & $\le-20.0$  &  $-20.44$  & $0.7-1.0$   & $0.87$ & $3.63 \pm0.18$ & $1.84 \pm0.12$ & $1.29 \pm0.20$ & $270 \pm50$ \\ %
Blue: Bluer &  1214         & $1.1 \ 10^{-3}$ & $0.56$          & $\le-20.0$  &  $-20.33$  & $0.7-1.0$   & $0.86$ & $3.16 \pm0.15$ & $1.78 \pm0.14$ & $1.11 \pm0.28$ & $260 \pm70$ \\ %
MR+SAM Red\tablenotemark{b}: &  7875 & $1.4 \ 10^{-3}$ & $1.14$ & $\le-20.0$  &  $-20.40$  & $0.7-1.0$   & $0.89$ & $6.31 \pm0.25$ & $1.91 \pm0.04$ & $1.96 \pm0.14$ & $490 \pm30$ \\ %
MR+SAM Blue: & 23645        & $4.2 \ 10^{-3}$ & $0.65$          & $\le-20.0$  &  $-20.43$  & $0.7-1.0$   & $0.88$ & $3.27 \pm0.07$ & $1.50 \pm0.03$ & $1.20 \pm0.10$ & $140 \pm20$ \\ %
\enddata
\tablenotetext{a}{For a \lcdm model with $\sigma_8=0.9$.}
\tablenotetext{b}{MR+SAM: Millenium Run with semi-analytic model}
\label{samples}
\end{deluxetable}

\clearpage

\clearpage

\end{document}